\documentclass[11pt,a4paper]{article} 
\usepackage{mathtools}
\usepackage[utf8x]{inputenc}     
\usepackage{jheppub}
\usepackage{enumerate}
\usepackage{epsfig} 
\usepackage{float} 
\usepackage{booktabs}
\usepackage[caption = false]{subfig}
\usepackage{wasysym} 
\usepackage{mathrsfs} 
\usepackage{amsfonts} 
\usepackage{amsbsy} 
\usepackage{amscd} 
\usepackage{pstricks} 
\usepackage{multirow} 
\usepackage{tikz}
\usepackage{color}
\usepackage{slashed}
\usepackage{array}
\usepackage{tablefootnote}

\usetikzlibrary{arrows,positioning,shapes.geometric} 
\usepackage[compat=1.1.0]{tikz-feynman}          
\usepackage[font=small,labelfont=bf]{caption}
\usepackage{slashed}
\usepackage{multirow}
\usepackage{tabularx}
\usepackage{soul}  
\usepackage{listings} 
\usepackage{color}

\usepackage{amsthm}

\newtheorem*{definition*}{Strong Inductive Bias}

\newcommand{\im}{\text{Im}}

\newcommand{\GL}{\text{GL}}

\title{

Optimal Equivariant Architectures from the Symmetries of Matrix-Element Likelihoods
} 
	\author{Daniel Ma\^\i{}tre,} 
\author{Vishal~S.~Ngairangbam,}
\author{and Michael~Spannowsky} 
\affiliation{Institute for Particle Physics Phenomenology, Department of Physics \\Durham University, Durham DH1 3LE, United Kingdom}
\emailAdd{daniel.maitre@durham.ac.uk}
\emailAdd{vishal.s.ngairangbam@durham.ac.uk}
\emailAdd{michael.spannowsky@durham.ac.uk}

\abstract{
	The Matrix-Element Method (MEM) has long been a cornerstone of data analysis in high-energy physics. It leverages theoretical knowledge of parton-level processes and symmetries to evaluate the likelihood of observed events. In parallel, the advent of geometric deep learning has enabled neural network architectures that incorporate known symmetries directly into their design, leading to more efficient learning. 
 This paper presents a novel approach that combines MEM-inspired symmetry considerations with equivariant neural network design for particle physics analysis. Even though Lorentz invariance and permutation invariance overall reconstructed objects are the largest and most natural symmetry in the input domain, we find that they are sub-optimal in most practical search scenarios. We propose a longitudinal boost-equivariant message-passing neural network architecture that preserves relevant discrete symmetries. We present numerical studies demonstrating MEM-inspired architectures achieve new state-of-the-art performance in distinguishing di-Higgs decays to four bottom quarks from the QCD background, with enhanced sample and parameter efficiencies. This synergy between MEM and equivariant deep learning opens new directions for physics-informed architecture design, promising more powerful tools for probing physics beyond the Standard Model.
}

\preprint{IPPP/24/69}

\allowdisplaybreaks
\begin{document} 
	\maketitle

\flushbottom	
\section{Introduction}

The search for new physics at the Large Hadron Collider (LHC) is a complex and data-intensive challenge. As particle collisions produce high-dimensional data, distinguishing between Standard Model events and potential new physics requires sophisticated analysis techniques. Traditionally, matrix-element methods (MEM)~\cite{Kondo:1988yd,Kondo:1991dw,D0:2015cyk,Alwall:2010cq,Soper:2011cr,Andersen:2012kn,Campbell:2012cz,Debnath:2014eaa,Soper:2014rya,Martini:2015fsa,FerreiradeLima:2016gcz,Prestel:2019neg,Martini:2020rnb,Bury:2020ewi,Butter:2022vkj,Grossi:2023fqq,Heimel:2023mvw} have been used to compare observed data to theoretical predictions by evaluating the likelihood of various hypothesized processes. In parallel, the advent of deep learning has enabled the development of powerful algorithms capable of learning complex patterns in data~\cite{deOliveira:2015xxd,Cranmer:2015bka,Dery:2017fap,Metodiev:2017vrx,Larkoski:2017jix,Komiske:2018cqr,Brehmer:2018eca,Guest:2018yhq,Qu:2019gqs,Brehmer:2019xox,Kasieczka:2019dbj,Karagiorgi:2021ngt,Plehn:2022ftl,Onyisi:2022hdh,doi:10.1142/12200,Brehmer2021,Maitre:2021uaa,DeZoort:2023vrm,Ngairangbam:2023cps,Bhardwaj:2024djv,Fenton:2020woz,Shmakov:2021qdz,Fenton:2023ikr,Chiang:2024pho,Maitre:2022xle,Rizvi:2023mws,Janssen:2023ahv,Bahl:2024meb,Bhardwaj:2024wrf}, often outperforming conventional methods in classification tasks.

In recent years, geometric deep learning~\cite{7974879,pmlr-v48-cohenc16,pmlr-v80-kondor18a,pmlr-v97-cohen19d,bronstein2021geometric,pmlr-v139-satorras21a,villar2021scalars,NEURIPS2023_6f6dd92b}  has emerged as a promising framework for physics analysis, incorporating known symmetries of physical laws directly into the neural network architecture. This approach, which could be called equivariant neural network design, seeks to restrict the learning task to a smaller yet appropriate class of functions by embedding symmetries such as Lorentz and permutation invariances into the model structure~\cite{pmlr-v119-bogatskiy20a,Bogatskiy:2022hub,Gong:2022lye,Li:2022xfc,Bogatskiy:2023nnw,Hao:2022zns,Thais:2023hmb,Spinner:2024hjm}. The general intuition that guides such architecture design is the invariance of physical observables under group transformations.

Despite the natural synergy between MEM, which explicitly utilises theoretical knowledge of symmetries through matrix element calculations and equivariant neural networks, a systematic connection between these two approaches has not been fully established. This work aims to bridge this gap by demonstrating how MEM-inspired symmetries can guide the design of equivariant neural network architectures for event classification tasks at the LHC. When deciding on which symmetries to embed in the model, we will show that the considerations that should guide the choice are the symmetry of the \emph{target function} rather than the physical symmetries of the network input. For example, in the case we consider here, the symmetry to use is that of the likelihood ratio and not the full Lorentz invariance of the input momenta.  We highlight the benefits of embedding optimal symmetries derived from the matrix-element calculations into the neural network to enhance classification accuracy, sample efficiency, and generalisation capabilities.

\begin{figure}
    \centering 
        \includegraphics[scale=0.4]{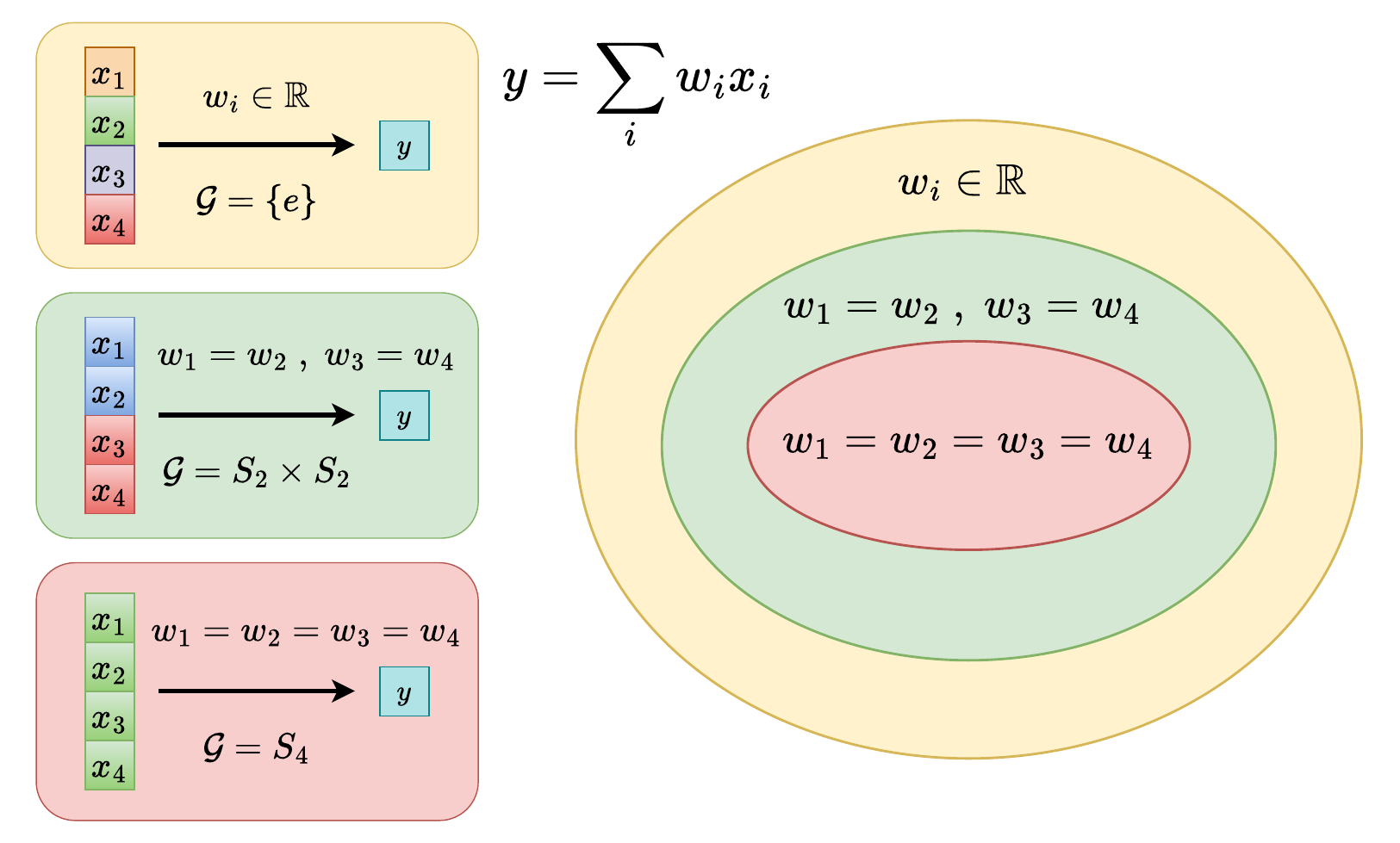}
            \caption{Representing the hierarchy in group invariant function approximation where a larger group ($S_4$) imposes additional constraints on the weights compared to a proper subgroup ($S_2\times S_2$). Although the constraints of $S_2\times S_2$ can become those of $S_4$, the stronger constraints of $S_4$ cannot become a function that is  $S_2\times S_2$ invariant but not $S_4$ invariant, as its weights lie strictly outside the red ellipse with the constraint $w_1=w_2\neq w_3=w_4$. Therefore, even though $S_4$ contains the group $S_2\times S_2$, an $S_4$-invariant function cannot become a purely $S_2\times S_2$-invariant function. This holds for general group invariant functions due to the structure of fibres induced by invariance in the function's domain (see fig.~\ref{fig:inv_fibre}). 
            }
    \label{fig:inv_comp}
\end{figure}

We begin by discussing the role of symmetries in function approximation, where they manifest themselves as group orbits in the equivalence classes of a target function's fibre in Section~\ref{sec:heur_induct}. We then explore optimal symmetry group choices for classification problems using the Neyman-Pearson lemma and their connection to the fibres of group-equivariant functions in Section~\ref{sec:opt_sym}. While the arguments based on group actions are more general, a simplified example which helps explain the hierarchy of group invariant function approximation is shown in fig.~\ref{fig:inv_comp}.  The general intuition can be stated as follows:

\begin{quote} 
In signal vs. background classification tasks, one can infer the (approximate) symmetries of the target function (not the data) from the specific processes' underlying likelihood based on the differential cross-sections. A universally approximating equivariant architecture on the space of functions with smaller or the same symmetries can approximate the target function but not one with a strictly larger symmetry. 
\end{quote} 
Incorporating Lorentz symmetry and permutation invariance, essential in evaluating cross-sections, provides a foundation for developing equivariant architectures. Building on this theoretical groundwork,  we investigate the optimality of the Lorentz invariance and $S_n$ permutation invariance over all  $n$ reconstructed objects for the evaluated MEM-likelihoods in Section~\ref{sec:equi_mem}. The former is suboptimal due to the dependence of the event likelihood on the transfer function, which is invariant only under longitudinal boosts and rotations along the $z$-axis.  The $S_n$ group is optimal when the final state consists of a single type of reconstructed object.\footnote{Any event-level analyses on reconstructed objects with point cloud architectures which assume $S_n$ invariance is, therefore, suboptimal in the sense of Neyman-Pearson when there is more than one type of reconstructed object. While their good performance may be due to the negligible null orbits of finite group symmetries in an uncountable domain, even in this suboptimal situation, they mostly outperform shallow machine learning on high-level features, which is a testament to the expressibility of modern deep learning algorithms.} Therefore, we devise a longitudinal boost invariant homogeneous\footnote{The requirement of permutation symmetry under separate classes of reconstructed objects allows for a heterogeneous graph construction. We do not consider such an approach as it has a factorial growth of learnable functions based on the number of edge and node types.} message passing neural network, where the smaller permutation symmetries are maintained by concatenated sub-graph readouts.  

To illustrate the practical implications of our approach, in Section~\ref{sec:dihiggs}, we present a case study of di-Higgs production with four bottom jets in the final state, a channel of particular interest for probing the Higgs self-coupling at the LHC. We demonstrate that MEM-inspired symmetries improve network performance in classification metrics compared to state-of-the-art results~\cite{Chiang:2024pho} and maintain better performance metrics with up to three orders of magnitude fewer parameters. 

Our findings suggest that by integrating the principles of the matrix-element method with modern equivariant deep learning techniques, we can develop more efficient and physically informed architectures for LHC data analysis. This synergy paves the way for new methodologies in the ongoing search for physics beyond the Standard Model.

\section{Symmetries as Strong Inductive Biases} 
\label{sec:heur_induct}
The theoretical reasoning behind symmetries becomes evident from its relation to conserved quantities, i.e. a symmetry transformation on a physical system does not change observable quantities. This carries over to function approximation, as the value of physically meaningful functions should not change under a symmetry transformation in the input feature space. Even without such symmetry considerations, defining a function requires each element on its domain to be associated with only one element in its co-domain (not one-to-many). Therefore, any given function divides the domain into mutually exclusive subsets mapped to the same element on its co-domain. These subsets called the function's fibres, are a particular \emph{partition} of the input domain unique to a family of functions.

\begin{figure}
    \centering 
        \includegraphics[width=\textwidth]{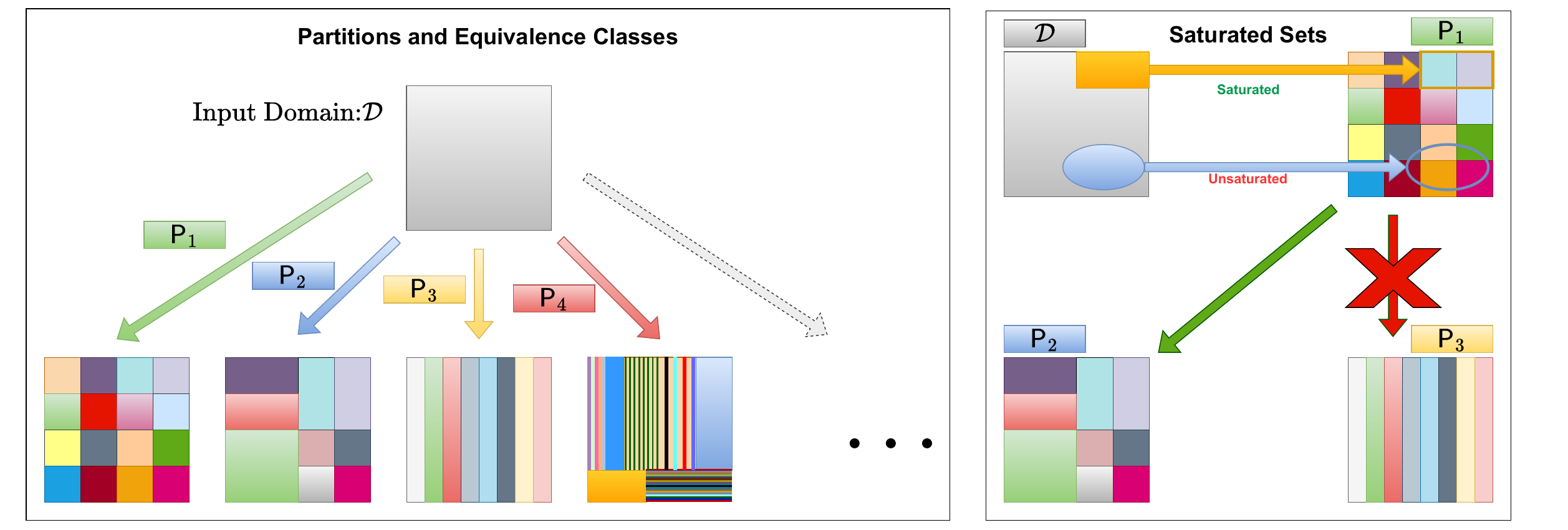}
            \caption{The left shows some possible partitions of a bounded domain $\mathcal{D}$ in 2D out of infinitely many possibilities. On the top right, the yellow rectangle is a saturated set in $\mathsf{P}_1$ while the blue ellipse is not. Consequently, if one restricts the smallest possible fibres that a function approximator can have to be those in $\mathsf{P}_1$, it can accommodate a target function with $\mathsf{P}_2$ (bottom left of figure on the right) as its fibres since all partitions in  $\mathsf{P}_2$ are saturated under $\mathsf{P}_1$. However, if it had $\mathsf{P}_3$, no amount of function approximation on $\mathsf{P}_1$ will agree over the whole domain $\mathcal{D}$ since all of its fibres are unsaturated in $\mathsf{P}_1$. Note that for incompatibility, one fibre being unsaturated is sufficient for incorrectness of $\mathsf{P}_1$. 
            }
    \label{fig:inf_parts}
\end{figure}

A partition of a set is a collection of subsets that do not have any element in common and, together, contain all elements of the parent set. Each set in this collection forms an \emph{equivalence class} in the set we refer to as \emph{blocks}. There are infinitely many ways of constructing such partitions of a set with infinite elements. These are diagrammatically illustrated on the left in fig.~\ref{fig:inf_parts}. If a subset of the parent set can be expressed as a union of blocks of a partition, this subset is said to be \emph{saturated} in the said partition. If the subset has a non-empty intersection with a block but without containing all of its contents, it is called \emph{unsaturated}. For example, on the top right in fig.~\ref{fig:inf_parts}, the yellow rectangle is saturated in the partitions of $\mathsf{P}_1$, while the blue ellipse is unsaturated.

In function approximation, the target function's fibre corresponds to a unique partition that corresponds out of all possible partitions of the input feature space. Therefore, the process of function approximation can be broken down into two stages:
\begin{enumerate}
\item finding a partition on the domain which matches that of the target function
\item matching the target function's value on these domain partitions over the family of functions having the same fibres
\end{enumerate} By strong inductive biases, we mean the assumption of a partition on the domain, which helps in the first stage and is related to the specifics of the data representation and its associated architecture. The second part is related to the actual function finding via an optimisation algorithm where one can include additional information as weaker forms of inductive biases without hard restrictions on the partitions. For instance, regularisation terms on the loss function will prioritise a region of the weight space without a hard boundary. While our definition can be made more general to encompass such biases, we do not consider such a generalisation since the former is a necessary condition for the latter and is the primary motif of the work. 

One can now define a strong inductive bias in terms of assumed partitions on the domain:
 \begin{definition*}
  	Given an approximation problem where we want to learn a continuous target function $f:\mathcal{D}\to\mathcal{H}$ between the domain $\mathcal{D}$ and the co-domain $\mathcal{H}$ via an approximator $\hat{f}:\mathcal{D}\to\mathcal{H}$ belonging to a family of functions $\Sigma$, a strong inductive bias is an assumption of a partition of the domain $\mathcal{D}$, such that $\hat{f}(\mathbf{x})$ has constant value in each block of the partition for all $\hat{f}\in\Sigma$.    
  \end{definition*}  
  \noindent 
  Let the partitions be represented as $\mathsf{\hat{P}}=\{[\mathbf{x}]^a_{\Sigma} :a\in I\}$, with $I$ being a set which indexes each block $[\mathbf{x}]^a_{\Sigma}$. 
  Since the collection $\mathsf{\hat{P}}$ is a partition of the domain, $\mathcal{D}=\bigcup_{a\in I} [\mathbf{x}]^a_{\Sigma}$ , and $[\mathbf{x}]^a_{\Sigma}\cap [\mathbf{x}]^b_{\Sigma}=\emptyset$ for $a\neq b$ and $[\mathbf{x}]^a_{\Sigma}=[\mathbf{x}]^b_{\Sigma}$ otherwise. The assumed partitions define the smallest mutually exclusive subsets of the domain, where an approximated function should be equal. Therefore, a strong inductive bias defines a function space on the domain where any function's value has to be constant within a single block while they can be different in separate blocks as a whole. Additionally, there is no restriction to the functions becoming equal in two distinct blocks. Therefore, the partitions $\mathsf{\hat{P}}$ are the \emph{minimal fibres} over the function space $\Sigma$.

Given the input feature space, the assumption of a partition reduces the learning process (the optimisation stage) to learning over single representatives from the equivalence classes. While it is most straightforward to encode the target function's fibre as partitions of the domain,  their exact fibres are never known in practice. As a result, partitioning the domain to help achieve the target function's fibre demands a notion of compatibility. For a given target function $f:\mathcal{D}\to\mathcal{H}$, it essentially boils down to the comparison of two partitions in $\mathcal{D}$: 
\begin{itemize}
\item The partitions $\mathsf{P}$ induced by the target function's fibres, say $[\mathbf{x}]_f$, where the function is equal in each block $[\mathbf{x}]_f\in\mathsf{P}$
\item The smallest possible fibres restricted via the inductive bias in all functions $\hat{f}:\mathcal{D}\to\mathcal{H}$ represented by the neural network architecture class, say $\mathsf{\hat{P}}\ni[\mathbf{x}]_{\Sigma}$. 
\end{itemize}

In the sense of an exact representation,\footnote{The case involving an $\epsilon$-accurate approximation with $\epsilon>0$ is more involved and will be touched upon in a future work~\cite{induct}. For the present work, it suffices to regard that the exact representation belongs to the restricted function space where the relevant architecture class with an inductive bias have (or should have) the universal approximation property.} the requirement is that a strong inductive bias (or partitions of the domain) is compatible with a target function if all of its fibres are saturated sets in the assumed partitions i.e. 
$$[\mathbf{x}]^b_f=\bigcup_{a\in I_b}\,[\mathbf{x}]^a_{\Sigma}\quad$$ for every $[\mathbf{x}]^b_f$ in $\mathsf{P}$, with $I_b$ an index set for each $[\mathbf{x}]^b_f$. 
If this does not hold true, the target function has two distinct fibres in some partition $[\mathbf{x}]_{\Sigma}$ in $\mathsf{\hat{P}}$ and any $\hat{f}$ cannot simultaneously become equal to both values in $[\mathbf{x}]_{\Sigma}$. Going back to the bottom right of fig.~\ref{fig:inf_parts}, an inductive bias of $\mathsf{P}_1$ is correct if the target function has fibres that correspond to $\mathsf{P}_2$, and incorrect if its fibres are $\mathsf{P}_3$. 

In particle physics applications, target functions are generally invariant under a group and therefore, elements belonging to each partition $[\mathbf{x}]_f$ are related by symmetry transformations. Due to the nice algebraic properties of elements in each fibre of the target function, it is comparatively straightforward to construct architectures which respect these symmetries. Therefore, symmetries play an important role in function approximation tasks. As we shall see, the main difference to the usual notion of symmetries is that the largest possible physical symmetry in the input domain is not necessarily the best choice since it enlarges the minimal fibres compared to its subgroup symmetries.

\section{Optimal Symmetries in Group Invariant Classification} 
\label{sec:opt_sym} 

For a group $\mathcal{G}$ with corresponding transformations $\rho_\mathcal{D}$ and $\rho_\mathcal{H}$ on the domain and co-domain, respectively, a function $f:\mathcal{D}\to\mathcal{H}$ is equivariant with respect to these transformations if 
\begin{equation}
    \label{eq:equi_def}
    f(\rho_\mathcal{D}(g)\,\mathbf{x})=\rho_\mathcal{H}(g)\,f(\mathbf{x})\quad.
\end{equation}
If the representation $\rho(g)_\mathcal{H}$ is trivial ($\rho_\mathcal{H}(g)=\mathbf{1}\,\forall g\in\mathcal{G}$), then $f$ is called $\mathcal{G}$-invariant. 
Particle fields in Quantum Field Theory are classified in terms of their transformation properties under the Lorentz group, and interacting theories are written down with Lorentz invariant Lagrangians and additional internal symmetries. This is the origin of the symmetries of the differential cross-section. For instance, take the transformation
$$\psi(\Lambda(g) \;p_\nu)=W(g)\;\psi(p_\nu)\quad,$$  of the Weyl spinor $\psi(p_\mu)$, $\psi:\mathbb{R}^4\to\mathbb{C}^2$, under a Lorentz group element $g$, where $W(g)$ and $\Lambda(g)$, respectively, are the Weyl and vector representations of the Lorentz group. While the term ``equivariance" is seldom used in QFT textbooks, classically, $\psi$ is a Lorentz equivariant function under the defined group transformations.

The nature of perturbative differential cross-sections already contains a rich structure of symmetries without going into the specific details of processes. On the other hand, the search for new physics is essentially a hypothesis test with the null background-only hypothesis vs the alternate signal and background hypothesis. With optimality of the likelihood ratio, guaranteed by the Neyman Pearson lemma~\cite{Neyman:1933wgr}, one can study the optimality of an imposed group equivariance by checking whether the space of a family of group equivariant functions contains monotonic functions of the likelihood ratio.  We briefly discuss this connection by describing the structure of fibres of group equivariant functions and its relation to the Neyman-Pearson optimality of group invariant likelihood ratios. This is essentially a condensed summary of ref~\cite{Ngairangbam:2024cxc}.

\subsection{Equivariant function spaces}
\label{sec:equi_level} 
As mentioned in fig~\ref{fig:inv_comp}, the underlying motivation for choosing correct symmetries is the comparable constraints of a hierarchical set of group invariant functions. More precisely, there is a set-inclusion relationship within the space of invariant functions of a group and its subgroup, which goes in the opposite direction of group inclusions. Take a group  $\mathcal{G}_1$ and its proper sub-group $\mathcal{G}_2$, i.e.  $\mathcal{G}_2\subsetneq\mathcal{G}_1$. On the same domain and a given group action of the group $\mathcal{G}_1$, restricting the group elements to those in $\mathcal{G}_2$ creates a $\mathcal{G}_2$-action. For group invariance (i.e. trivial action on the co-domain), this creates two invariant function spaces on the domain $\mathcal{D}$: say $\mathcal{F}_{\mathcal{G}_1}$ and $\mathcal{F}_{\mathcal{G}_2}$. Since all $\mathcal{G}_1$ invariant functions are $\mathcal{G}_2$ invariant, but not all $\mathcal{G}_2$ invariant functions are $\mathcal{G}_1$ invariant, we have : $\mathcal{F}_{\mathcal{G}_2}\supsetneq\mathcal{F}_{\mathcal{G}_1}$. This means that functions which are $\mathcal{G}_2$ invariant but not $\mathcal{G}_1$ invariant do not belong to $\mathcal{F}_{\mathcal{G}_1}$. A schematic diagram depicting this inverted hierarchy in the function space is shown in fig.\ref{fig:inv_fibre}. Assuming the target function is always group invariant, a qualitative explanation of why this happens is given separately for the $\mathcal{G}$-invariant classification and $\mathcal{G}$-equivariant feature extraction.

\subsubsection*{$\mathcal{G}$-invariance}
\begin{figure}
    \centering 
        \includegraphics[scale=0.4]{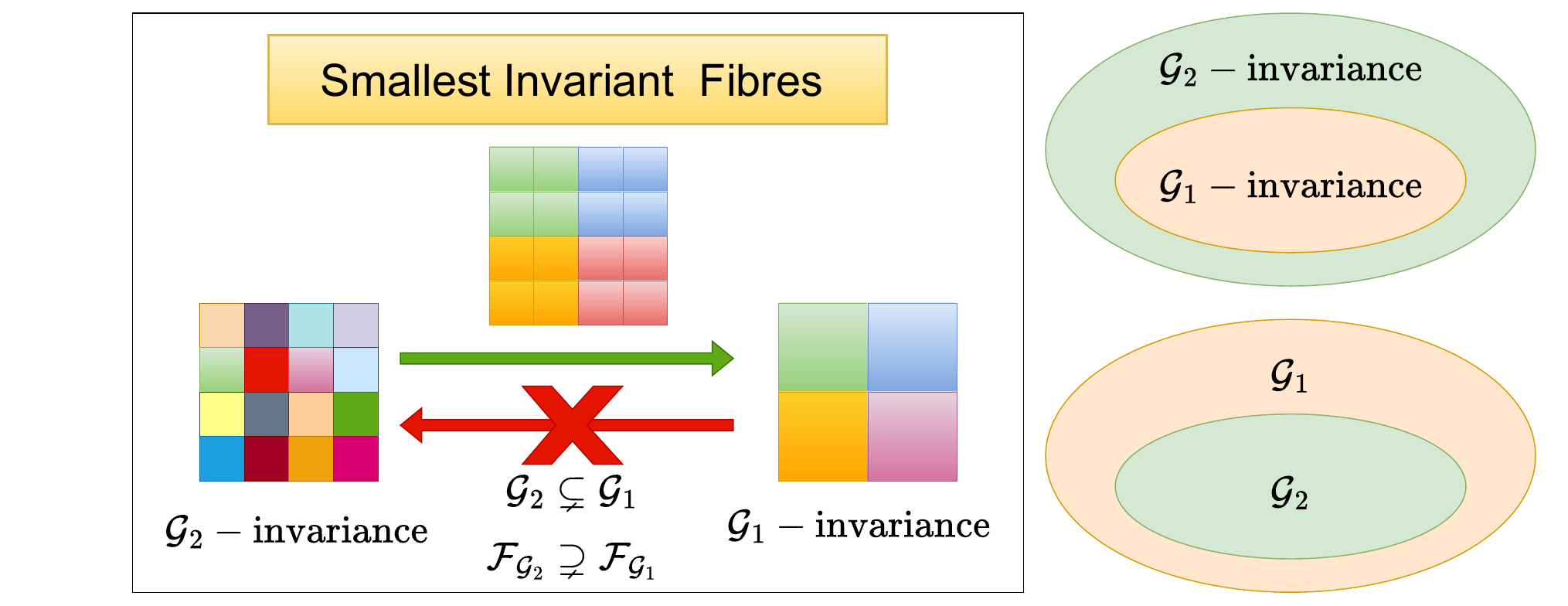}
            \caption{ The difference between the smallest \emph{fibres} (the set of points in the domain where the function's value is equal) of a function invariant under a group $\mathcal{G}_1$ and its proper subgroup $\mathcal{G}_2$. The smaller squares denote the coarsening of the domain with similar colours, signifying equality of the function's value.
            $\mathcal{G}_1$-invariance assumes larger fibres from the start. In contrast, $\mathcal{G}_2$-invariance assumes smaller and \emph{compatible partitions} with $\mathcal{G}_1$-invariance, i.e. they can become enlarged so that the function becomes equal on the smallest fibres of $\mathcal{G}_1$-invariant functions.  
            } 
    \label{fig:inv_fibre}
\end{figure}

Suppose a given function $f:\mathcal{D}\to\mathcal{H}$ is invariant under a transformation $\rho_\mathcal{D}(g)$ of a group $\mathcal{G}$. This means that $f(\rho_\mathcal{D}(g)\,\mathbf{x})=f(\mathbf{x})\equiv\mathbf{y}$ for all $g\in\mathcal{G}$, i.e. the fibre of an element $\mathbf{y}$ in the image of the function $\im(f)$, is at least as large as all those elements which can 
be traversed from $\mathbf{x}$ via the group action $\rho(g)_\mathcal{D}\,\mathbf{x}$. This subset of elements in $\mathcal{D}$ is the orbit of $\mathbf{x}$ under the $\mathcal{G}$-action.   While a like-for-like comparison between different group invariant neural networks is highly non-trivial, the structure of the smallest fibres (see fig.~\ref{fig:inv_fibre}) induced by group invariance in the input domain $\mathcal{D}$ provide a mathematically consistent mechanism of checking the suitability of a particular group invariance in the input domain even without recourse to the specific detail of the architecture or the universal approximation property. The important observation which allows such an inspection is that restricting the group action on the domain $\mathcal{D}$ to group elements of a proper subgroup generally\footnote{Mathematically, the group action should be \emph{effective} in that any non-identity group element has at least one non-trivial action on an element of the domain. This property is generally satisfied by group actions utilised in particle physics.} results in smaller minimal fibres as they have smaller orbits. Crucially, larger group invariance forces a function to be equal in different orbits of the subgroup action and is, therefore, not a correct symmetry when the target function is invariant only under a proper subgroup but not under the parent group. On the other hand, since group invariance only fixes the smallest fibres of a function, an expressive enough invariant network of a smaller group can approximate a function invariant under a larger group. The state-of-the-art performance of transformers~\cite{NIPS2017_3f5ee243} for jet-tagging~\cite{pmlr-v162-qu22b} which match or outperform equivariant ones~\cite{pmlr-v119-bogatskiy20a,Gong:2022lye,Bogatskiy:2023nnw,Spinner:2024hjm} is an extreme example of an architecture learning the relevant fibre structure of the target function without continuous group algebraic constraints in the domain. A diagrammatic representation of the compatibility of a subgroup invariance for a target function invariant under a larger group and incompatibility of a larger group invariance for a proper subgroup invariant target function is shown in fig~\ref{fig:inv_fibre}.

\subsubsection*{$\mathcal{G}$-equivariance}
Now consider that the function $f:\mathcal{D}\to\mathcal{H}$ is equivariant with respect to is a corresponding non-trivial transformation $\rho_\mathcal{H}(g)$ of the group acting on the co-domain, i.e. $f(\rho_\mathcal{D}(g)\,\mathbf{x})=\rho_\mathcal{H}(g)\,f(\mathbf{x})$. In such a case, the function is equal for at least those elements $\mathbf{x}'=\rho_\mathcal{D}(g)\,\mathbf{x}$ in the domain $\mathcal{D}$, transformed by group elements $g$ which fixes $\mathbf{y}=f(\mathbf{x})$, i.e. $\mathbf{y}=\rho_\mathcal{H}(g)\;\mathbf{y}$. This subgroup of $\mathcal{G}$ is dependent on the representation $\rho_\mathcal{H}(g)$, and the particular element $\mathbf{y}$ is the little group~\cite{wigner_little_group} of the group transformation for $\mathbf{y}$. For group invariant binary classification of signal and background events, one can consider that $\mathcal{H}$ is a hidden representation where we extract the relevant features as the target function. Within this, there are two extremes depending on the nature of the representation $\rho_\mathcal{H}(g)$ in the co-domain $\mathcal{H}$. If the action is \emph{free}, i.e. the little group of every element in $\mathcal{H}$ is the trivial group consisting only of the identity, equivariant feature extraction does not assume any larger fibres than the one assumed by an invertible function between the input domains. Therefore, for any noticeable gain in inductive biases, the group action on $\mathcal{H}$ should not be free. At the other extreme, if the little group of all elements in $\mathcal{H}$ is the group itself, then $f$ is $\mathcal{G}$-invariant. Therefore, in the case of group equivariant feature extraction for an invariant target function, the little group of all the elements in the co-domain should be no larger than the largest subgroup under which the target function is invariant.  One should remember that our discussions relate to the equivariant approximation of an invariant target function. For equivariant target functions, the purpose of equivariance beyond the assumption of a fibre structure is an efficient generalisation of unseen input data related via group transformations. Here, a correct free group action on the co-domain will offer advantages compared to non-equivariant ones in generalisation capabilities. Moreover, given a free group action on the co-domain, one can build subgroup invariants out of the equivariant quantities, manually inducing appropriate little groups. Such an approach would be suitable, for instance, in multi-class classification tasks where the different likelihood ratios are invariant under different subgroups of a parent group.

\subsection{Neyman-Pearson optimality and group equivariance}
\label{sec:neyman_pearson} 
Consider the binary classification problem of a signal hypothesis $\mathcal{P}_S$ with the corresponding set of processes $\mathcal{P}_B$ from the known sector of the Standard Model forming the background hypothesis. Each hypothesis $H\in\{S,B\}$  has a normalised probability densities $p_H(\mathbf{E})=\frac{1}{\sigma_{H}}\frac{d\sigma_H}{d\mathbf{E}}$, with $d\sigma_H$ and $\sigma_H$, the differential and integrated cross-section for the set of processes $\mathcal{P}_H$. From the Neyman-Pearson lemma, an optimal classifier between the two hypotheses is a monotonic function of the likelihood ratio\footnote{Generally, the alternate hypothesis for a signal search at the LHC is the presence of both signal and background processes, in which case the probability distribution is $p_{1}(\mathbf{E})=\frac{1}{\sigma_S+\sigma_B}(\frac{d\sigma_S}{d\mathbf{E}}+\frac{d\sigma_B}{d\mathbf{E}})$. Therefore, the likelihood ratio is $\lambda(\mathbf{E})=\frac{\sigma_B}{\sigma_S+\sigma_B}(1+\frac{d\sigma_S/d\mathbf{E}}{d\sigma_B/d\mathbf{E}})$. Our case considers the behaviour of the non-constant second term as the symmetry properties depend on this term alone. } $\lambda(\mathbf{E})=p_S(\mathbf{E})/p_B(\mathbf{E})$. Thereby, for a group equivariant neural network to approximate a monotonic function of the likelihood ratio, the smallest fibres assumed via group equivariance should be comparable to that of the likelihood ratio. Recollecting the nature of group equivariant fibres discussed above, one can construct the following guidelines for an optimal choice of the group $\mathcal{\hat{G}}$ given a $\mathcal{G}$-invariant likelihood ratio:

\begin{itemize} 
	\item \textbf{$\mathcal{\hat{G}}$-invariance}: $\mathcal{\hat{G}}$ can be a subgroup of $\mathcal{G}$ but not larger 
	\item \textbf{$\mathcal{\hat{G}}$-equivariance}: The little group of the $\mathcal{\hat{G}}$-action on the co-domain should not be larger than $\mathcal{G}$. 
	\end{itemize}  
	For the $\mathcal{\hat{G}}$-equivariant case, a free action on the co-domain will be compatible with any target function. Still, it will not provide any noticeable gain in generalisation ability compared to non-equivariant architectures. These guidelines also hold for any general $\mathcal{G}$-invariant target function. 
	
	The guidelines provide little utility in binary classification tasks when one knows the group $\mathcal{G}$. The real utility of these guiding principles arises in  $\mathcal{\hat{G}}$-equivariant feature extraction for multi-class classification where each class $c$, has a possibly different $\mathcal{G}_c$-invariant probability distribution. One can then use knowledge of the invariant probabilities to identify the group invariant likelihood for one-vs-one and one-vs-many classification scenarios and construct a $\mathcal{\hat{G}}$-equivariant function, which contains all these possibilities as its subgroup and has a compatible action with the final sub-group invariances in the intermediate feature extraction layers. One can enforce the appropriate little group invariances of the different possibilities at the final feature extraction layer to feed into the classifier head. 
	
	In signal searches, different processes in $\mathcal{P}_H$ may have a different set of permutation symmetry. For instance, the event weight in the resonant decay of a $Z$ boson to a pair of leptons will be invariant under their exchange, while it will not be if they originate from a pair of $W^\pm$ bosons. Such physical arguments open up an avenue for the design of equivariant architectures tailored to particular search scenarios, which on top of theoretically\footnote{depending on the universal approximation property of the equivariant architecture class} being able to approximate a monotonic function of the likelihood ratio will have better parameter and sample efficiency. They can also be used to modify the architecture of foundation models before fine-tuning for particular search scenarios.

\section{Equivariant architectures from the matrix-element method}
\label{sec:equi_mem}
 In the point cloud representation, one regards the input as a set and learns a permutation-invariant function for all possible permutations of the elements. They generally utilise sum-decomposition in a latent space to account for variable cardinalities of the samples, which is known to have universally approximating properties as set~\cite{NIPS2017_f22e4747} and multi-set~\cite{pmlr-v97-wagstaff19a}  functions. 
However, to study the optimality of the permutation group action on the squared matrix elements, we will consider a point cloud sample as an ordered $n$-tuple where we define functions to be invariant under possibly different permutation groups $S_{n'}$, acting on $n'\leq n$ elements. In this section, we first discuss the Lorentz and permutation symmetries of fixed-order differential cross-sections. We then discuss optimal symmetries that are present in the matrix-element likelihoods and present a longitudinal boost equivariant architecture which respects these symmetries.  
\subsection{Symmetries in fixed-order differential cross sections}
\subsection*{Lorentz Symmetry} 
 Let $\mathbf{X}=(\mathbf{p}_1,\mathbf{p}_2,...,\mathbf{p}_n)$ be the four-vectors of a measured event at LHC. In addition to these four-vectors, we have a corresponding vector  $\mathbf{H}=(\mathbf{h}_1,\mathbf{h}_2,....,\mathbf{h}_n)$ containing additional information such as the type of the reconstructed object, flavour, charge etc. These properties determine the information available on the partonic process at reconstruction and the permutation symmetry of the differential cross-sections in addition to the quantum mechanical indistinguishability of identical particles. 
Representing the combined observed information of $\mathbf{X}$ and $\mathbf{H}$ as $\mathbf{E}=(\mathbf{p}_1\oplus\mathbf{h}_1,\mathbf{p}_2\oplus\mathbf{h}_2,...,\mathbf{p}_n\oplus\mathbf{h}_n)$, consider that there are $r$ incoherent but observationally identical (i.e. at reconstruction)  processes $\mathcal{P}=\{a_1\,b_1\to F_1,a_2\,b_2\to F_2,...,a_r\,b_r\to F_r\}$ that can lead to the production of this event. Here, $a_i$ and $b_i$ are the incoming partons, and $F_i$ represents the partonic final state. The leading-order differential cross section dependent on theory parameters $\theta$ can be written as  
\begin{equation}
	\label{eq:lhc_diff_cross}
 \begin{split} 
	d\sigma_\mathcal{P}(\mathbf{q}_1,\mathbf{q}_2,\mathbf{E},\theta) =\sum_{a_ib_i\to F_i\in\mathcal{P}} \int dx_{1}\;dx_{2}\;\frac{f_{a_i}(x_1)\; \,f_{b_i}(x_2)}{2 E_{cm}\,x_1x_2}&\;\delta^{(4)}(x_1\mathbf{q}_1+x_2\mathbf{q}_2-\sum_{j=1}^n \mathbf{p}_j)\;\\
 \times&|\mathcal{M}_i(x_1 \mathbf{q}_1,x_2\mathbf{q}_2,\mathbf{E},\theta)|^2\;d\Pi_n\quad, 
 \end{split} 
\end{equation}
where\footnote{We use the convention $\mathbf{p}=(p_x,p_y,p_z,E)$ for easier discussion of the transverse and longitudinal components in later sections.} $\mathbf{q}_1=(0,0,E_{cm}/2,E_{cm}/2)$ and $\mathbf{q}_2=(0,0,-E_{cm}/2,E_{cm}/2)$ are the incoming proton momenta with centre-of-mass energy $E_{cm}$, $|\mathcal{M}_i|^2$ is the Lorentz invariant squared matrix-element for the parton-level process $a_ib_i\to F_i$, $f_{a_i}$ and $f_{b_i}$ are the proton parton distribution functions of the parton $a_i$ and $b_{i}$, respectively,  and $d\Pi_n$ is the Lorentz invariant phase space (LIPS) of the $n$-body final state $d\Pi_n= \prod_{j=1}^n \frac{d^3p_j}{(2\pi)^32E_j}$.  Given a Lorentz group element $g$, the corresponding transformation of the final state $\mathbf{E}$ is 
\begin{equation} 
	\label{eq:lorentz_transform} 
	\Lambda_\mathbf{E}(g) \mathbf{E}=\left(\Lambda(g)\mathbf{p}_1\oplus\mathbf{h}_1,\Lambda(g)\mathbf{p}_2\oplus\mathbf{h}_2,...,\Lambda(g)\mathbf{p}_n\oplus\mathbf{h}_n\right)\quad,  
\end{equation} 
where the matrix representation $\Lambda_\mathbf{E}(g)$ can be built from the vector representation $\Lambda(g)$ acting on four vectors $\mathbf{p}_i$, and the trivial identity matrix representation acting on scalars $\mathbf{h}_i$. Events correspond to different points in the phase space whose relative weight is determined by the Lorentz invariant matrix-element squared $|\mathcal{M}_i|^2$, i.e. the probability distribution of a given final state signature under a hypothesised process $a_ib_i\to F_i$   is Lorentz invariant. 
At this point, the sum over all processes is also Lorentz invariant. However, experimental considerations render event likelihoods that do not respect the full Lorentz invariance. This will be discussed further in Section~\ref{sec:opt_sym_mem}.

\subsection*{Permutation Symmetries}  
Let the observed event be  $\mathbf{E}=(\mathbf{r}_1,\mathbf{r}_2,...,\mathbf{r}_n)$ such that $\mathbf{r}_i=\mathbf{p}_i\oplus\mathbf{h}_i$. The action of the $n$-object permutation group $S_n$ on $\mathbf{E}$, permutes each $\mathbf{r}_i$ as a whole
$$\rho(\sigma)(\mathbf{r}_1,\mathbf{r}_2,...,\mathbf{r}_n)=(\mathbf{r}_{\sigma(1)},\mathbf{r}_{\sigma(2)},...,\mathbf{r}_{\sigma(n)})\quad,$$  
where $\rho:S_n\to \GL(n\times (4+m),\mathbb{R})$ is a matrix representation of $S_n$ built as $\rho(\sigma)=\rho_n(\sigma)\otimes \mathbf{1}_{4+m}$,   out of the canonical representation $\rho_n(\sigma)$ of $S_n$ in $\GL(n,\mathbb{R})$, with $m$ being the dimensions of $\mathbf{h}_i$. Similarly for some $n'< n$, one can also define the permutation action on $n'$ elements  via a representation $\rho_{n'}: S_{n'}\to \GL(n,\mathbb{R})$ of $S_{n'}$ in $\GL(n,\mathbb{R})$.  Clearly, there are $\binom{n}{n'}$ ways of choosing subsets of cardinality $n'$ from $\mathbf{E}$, each having a particular form of the matrix $\rho_{n'}(\sigma')\in \GL(n,\mathbb{R})$, $\sigma'\in S_{n'}$ reflecting the chosen subset.   A function $f:\mathcal{E}\to\mathcal{H}$, where $\mathcal{E}$ is the space of measured events is permutation invariant if
\begin{equation}
\label{eq:perm_inv} 
    f(\rho(\sigma)(\mathbf{r}_1,\mathbf{r}_2,...,\mathbf{r}_n))=f(\mathbf{r}_1,\mathbf{r}_2,...,\mathbf{r}_n)\quad,
\end{equation}
for all $\sigma\in S_n$. The differential cross-section is not symmetric concerning the exchange of distinct particles, which results in the non-invariance of the likelihood under the exchange of reconstructed objects belonging to different classes. This will be discussed further in Section~\ref{sec:opt_sym_mem}.

\subsection{Optimal Symmetries from the matrix-element method} 
\label{sec:sym_mem} 
The matrix-element method is a theoretically motivated multivariate data analysis approach which evaluates the likelihood of an event arising from a set of parton-level processes $\mathcal{P}$. With a slight modification of eq~\ref{eq:lhc_diff_cross} to account for detector effects and implicitly considering momentum conservation, the likelihood of an event $\mathbf{E}$ arising due the $i$-th parton level process in $\mathcal{P}$ say $ab\to F$, is
\begin{equation}
	\label{eq:mem_weight} 
	p_i(\mathbf{E}|\theta)=\frac{1}{\sigma_i}\int d\Pi_n(\mathbf{P})\; dx_1\,dx_2\; \frac{f_a(x_1)f_b(x_2)}{2E_{cm}\,x_1x_2} |\mathcal{M}_i(x_1 \mathbf{q}_1,x_2\mathbf{q}_2,\mathbf{P},\theta)|^2\; T(\mathbf{E},\mathbf{P}) \quad.
\end{equation}
Here, $T(\mathbf{E},\mathbf{P})$ is the transfer function modelling the probability of the event $\mathbf{E}$ arising from the final state four-vectors $\mathbf{P}$ of the partonic configuration $F$. In conjunction with the integration over the parton-level LIPS $d\Pi_n(\mathbf{P})$, the transfer function accounts for detector effects which decide up to what extent the exact symmetries of $|\mathcal{M}_i|^2$ are carried over to the likelihood $p_i(\mathbf{E}|\theta)$ or add new discrete symmetries by making quantum mechanically non-identical partons indistinguishable due to experimental considerations.  The likelihood for the hypothesis set $\mathcal{P}_H$, with $\sigma_H=\sum_i\sigma_i$ is 
$$p_H(\mathbf{E}|\theta)=\frac{1}{\sigma_H}\sum_{i\in\mathcal{P}_H} \sigma_i\,p_i(\mathbf{E}|\theta)\quad.$$  
Therefore, in such a set-up, one can construct the likelihood and likelihood ratios of any set of non-interfering parton level processes. Moreover, for equivariant feature extraction, one can infer the (approximate) optimal group symmetries from each $p_i(\mathbf{E}|\theta)$.  

In this section, we highlight the general structure of symmetries inherent in the likelihoods while consistently taking resonant and non-resonant production of di-Higgs decaying to four bottom jets as an example to concretely illustrate the synergy between group equivariant architecture design and the probabilities $p_H(\mathbf{E})$. This is one of the most promising channels for looking into the quartic Higgs self-coupling at LHC, as it has the highest branching ratios but is plagued by a very high QCD multi-jet background, and we will consider it for the numerical analysis in the next section.

\label{sec:opt_sym_mem} 
\subsubsection*{Continuous symmetry}
In most searches, we are interested in a fixed number of primary partons: the four bottom quarks in the di-Higgs case. Due to the inevitability of additional QCD radiation at the very high energies of LHC, a rigid cut on the number of jets is sub-optimal as it throws away many possible signal events. To consider many events, one includes hard processes with additional QCD radiation beyond the four bottom quarks in the signal and the background sets of processes. Possibly coherent processes in the unresolved regime within these processes must be matched and merged with the parton-shower-generated additional radiation to avoid over-counting in the overlapping phase space regions.  Additionally, these processes involve a variable number of final state particles that do not live in the same phase space. Special care needs to be taken to evaluate such weights ~\cite{Alwall:2010cq,Soper:2011cr,Andersen:2012kn,Campbell:2012cz,Martini:2015fsa,FerreiradeLima:2016gcz,Prestel:2019neg}.  One mechanism is to introduce kinematic corrections on an event-by-event basis for manageable number of additional hard radiations,~\cite{Alwall:2010cq,Andersen:2012kn,Campbell:2012cz} so that the weights are evaluated in the phase space involving fixed number of primary partons. Such kinematic corrections are essentially a preprocessing stage in machine learning terminology.

\begin{figure}
    \centering
    \includegraphics[scale=0.5]{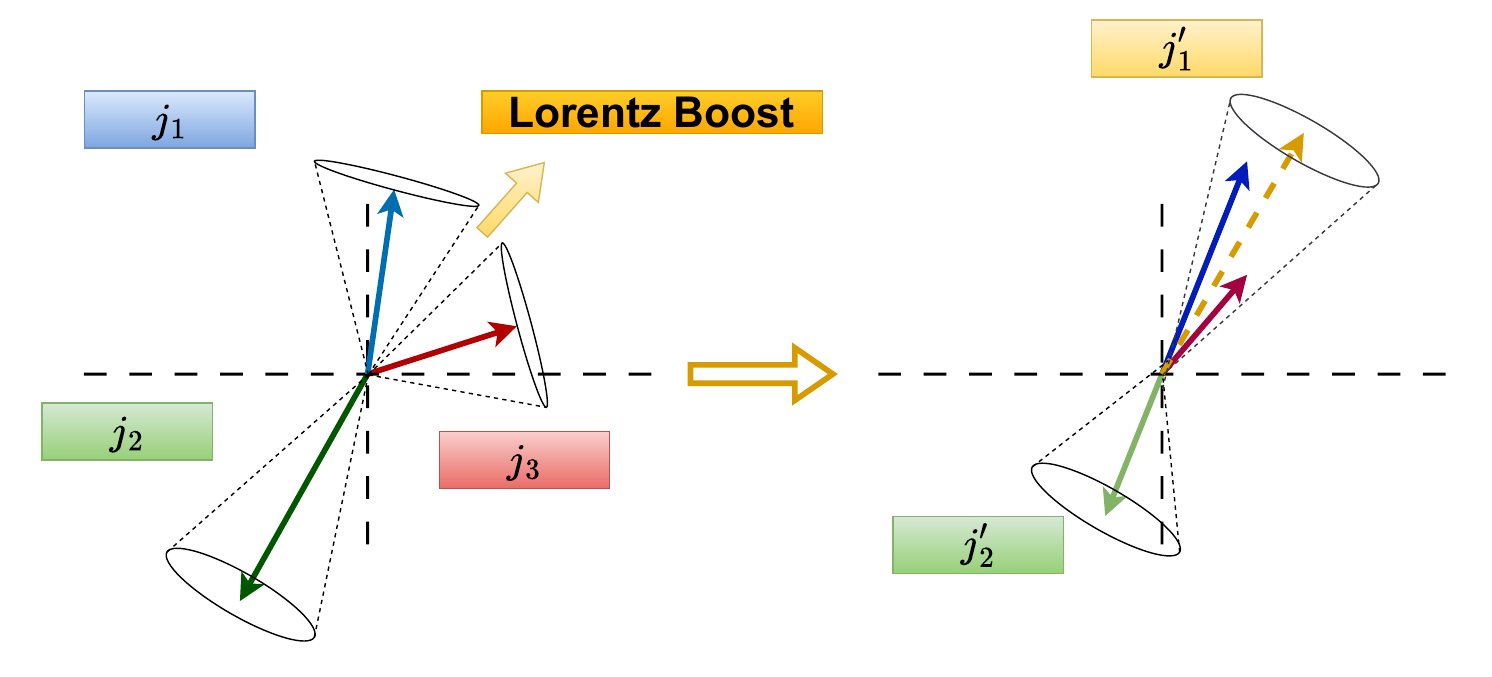}
    \caption{A Lorentz boost in the direction opposite to the leading jet in a three-jet event (on the left) will transform it into a two-jet event (on the right). For a baseline selection criteria which allows for more than two jets in the final state, the MEM-likelihood evaluated as a sum of three jet final state processes and two jet final state processes will not be the same for either event, making the event likelihood violate Lorentz invariance.}
    \label{fig:non_lorentz_inv}
\end{figure}

To bring in MEM-inspired symmetries into equivariant architecture design, we do not consider a kinematic preprocessing stage and consider the group invariance of 
MEM-weights of the sum over processes with a variable number of final state particles. In such a case, since the transfer function $T(\mathbf{E},\mathbf{P})$ involves the reconstruction algorithm and baseline selection criterion, the likelihood $p_i(\mathbf{E}|\theta)$ is not necessarily Lorentz invariant. For example, commonly used jet algorithms depending on $p_T$ and $\Delta R$ are longitudinal boost invariant but not fully Lorentz invariant. In fig.~\ref{fig:non_lorentz_inv}, we show a fully visible final state with three jets on the left, becoming a two-jet event on the right with an appropriate Lorentz boost. In the three jet event, the green leading jet has a large transverse momentum compared to the two sub-leading jets, a boost along the direction opposite to the leading jet will result in its momentum becoming lower with the two sub-leading jets coming closer. Once the sub-leading jets' angular separation is reduced to within the jet radius, the event will become a two-jet event, as shown on the right. The situation becomes more severe for signatures with invisible particles in the final state where there is no upper limit on the missing transverse momentum as there are many possible boost directions, which will result in two separated objects becoming unresolvable in the sample space of selected events since the momentum mismatch in the lab-frame will be regarded as belonging to the invisible particles and therefore belong to the sample space of selected events.

From this example, one can see that the event likelihood is not Lorentz invariant because of the non-invariance of the jet algorithm where the radius is kept fixed and the measures $\Delta R_{ij}$ transform non-trivially under general Lorentz boosts or rotations. The isolation criteria on other types of reconstructed objects and the jet algorithm are generally invariant under rotations, and Lorentz boosts along the $z$-axis, with the likelihood of maintaining invariance under such a sub-group. As a group which mixes the orbits under longitudinal boosts and rotations along the $z$-axis, the Lorentz group is strictly larger and, hence, an incorrect group.

\subsubsection*{Discrete Symmetry}
An event consists of sets of different reconstructed objects like leptons, light jets, bottom jets, photons, etc., which may be grouped into a single class or separated depending on the signal and background hypotheses. Denoting each object type as a vector $\mathbf{E}_\alpha$ with each $\alpha\in\{1,2,..,k\}$ specifying the object type of $k$ classes of reconstructed objects with cardinality $n_\alpha$, an event is represented as a vector\footnote{Strictly speaking, each $\mathbf{E}_\alpha$ as well as the full representation $\mathbf{E}$ is also a direct sum over $\mathbf{r}_i$. However, when considering the object properties, we will write all capital boldfaced vectors as a tuple of elements $\mathbf{r}_i$ to avoid confusion between the two situations.} $\mathbf{E}=\bigoplus_{\alpha}\mathbf{E}_\alpha$.  Since a $\mathcal{\hat{G}}$-invariant function approximator cannot efficiently approximate any $\mathcal{G}$-invariant functions when $\mathcal{G}$ is a proper subgroup, we need to determine the largest possible permutation symmetry of the likelihood and the likelihood ratios.  Again, this is entirely determined by 
 $T(\mathbf{E},\mathbf{P})$: for each reconstructed object $\mathbf{r}_i$ in $\mathbf{E}$, $T(\mathbf{E},\mathbf{P})$ assigns it all possible parton flavours within a sum. This renders $T(\mathbf{E},\mathbf{P})$ and hence $p_i(\mathbf{E}|\theta)$ invariant under the exchange of elements within the same reconstructed object class that have no charge information (i.e. jets, bottom-tagged jets, and photons but not electrons, muons, and tau jets).    
 Therefore, even if two particles are (considered) indistinguishable at reconstruction, they may be separate particles in the partonic final states like gluons and quarks. On the other hand, distinguishable particles at reconstruction are always non-identical at the parton level, and first-principle arguments do not guarantee permutation invariance of the matrix-element squared under their exchange in the final state. Therefore, \textit{if an observed event $\mathbf{E}$ with $n$ objects contains more than one reconstructed object type, or if it contains a single object type with at least two objects having different observed charges, the process likelihood $p_i(\mathbf{E}|\theta)$ is not $S_n$-invariant.} 

As a concrete example, 
let us consider a signature with two photons and three jets represented as $\mathbf{E}_\gamma=(\mathbf{r}^\gamma_1,\mathbf{r}^\gamma_2)$ and $\mathbf{E}_J=(\mathbf{r}^J_1,\mathbf{r}^J_2,\mathbf{r}^J_3)$. The largest symmetry in the underlying matrix elements is when all three jets originate from a gluon at the parton level. For this process, the matrix-element squared $|\mathcal{M}(\mathbf{r}^\gamma_1,\mathbf{r}^\gamma_2,\mathbf{r}^J_1,\mathbf{r}^J_2,\mathbf{r}^J_3)|^2$ is permutation invariant under the exchange of the two photons or within the exchange of gluons amongst themselves but not in the interchange of a photon and a gluon. Therefore, the MEM-likelihood is not $S_5$ permutation invariant. Almost all point cloud-based architectures studied for event-level analyses implicitly consider a full permutation invariant representation over the reconstructed objects regardless of the final state's composition. Even though this contains the smaller permutation symmetries of the MEM-likelihood ratio, $S_n$ permutation symmetry is a larger symmetry unless all reconstructed objects belong to the same type and are, therefore, not a correct symmetry for any given final state.

A straightforward solution which fixes the non-invariance of the target function under the exchange of elements belonging to different blocks in any point cloud approach, including graph neural networks, is to operate a sub-graph readout over the different classes $\mathbf{\hat{E}}_\alpha$ which segregates the reconstructed objects based on distinguishability and then concatenate these sub-graph representations. For instance, in a mean readout operation, the event representation   
\begin{equation}
	\label{eq:correct_reco_readout}
	\mathbf{\hat{E}}=\bigoplus_{\alpha=1}^k \left(\frac{1}{n_\alpha}\sum_{i=1}^{n_\alpha} \mathbf{\hat{r}}^\alpha_i\right)\quad, 
\end{equation}
fixes a particular ordering of the reconstructed object classes and is invariant only under permutations that act separately on each block vector $\mathbf{\hat{E}}_\alpha$'s constituents. So far, we have considered object reconstruction to have perfect accuracy. One should relax such rigid division of the reconstructed objects to account for experimental realities, including the possible absence of some classes in an event depending on the baseline selection criterion. This can be done by assigning relative weights  $w_{\alpha_1\leftarrow\alpha_2}\in(0,1)$ not necessarily symmetric, which controls the relative contribution of class $\alpha_2$ to the readout operation of $\alpha_1$. These weights could be learnt as an attention mechanism modified with the concatenation operation over the $\alpha_1$ axis. However, as proof of principle, we do not consider such modifications and set the weights beforehand in the architecture design for the numerical experiments. Even though the modified structure may not affect the performance of highly expressive networks, we speculate it will affect the theoretical uncertainties when merging additional radiations at higher perturbative accuracies. Since understanding such theoretical uncertainties is crucial in deploying deep learning algorithms for phenomenological studies, we leave an in-depth analysis of such an impact for independent future work.

\subsection{Approximate Symmetries under the Narrow Width Approximation}

As we have seen above, the largest permutation symmetry in an event is the product group $\otimes_{\alpha=1}^k S_{n_\alpha}$ permuting elements within the same class of reconstructed objects. 
 However, additional approximate symmetries may be smaller or larger depending on the process. For QCD background processes producing at least four bottom jets, the event weight is $S_4$ permutation invariant. In contrast, for the SM di-Higgs production within the narrow width approximation (NWA), out of the three possible partitions into two pairs of bottom quarks, the phase space volume where more than one of them lies near the mass peak is very small and hence, for most events, two out of the three distinct parton level pairings will have a negligible contribution to the overall event weight, giving us a reduced $S_2\times S_2$ approximate symmetry. On the other hand, if instead of the SM di-Higgs production, there is a resonant heavy Higgs with a very small width, the complete $S_4$ symmetry is approximately restored as the dominant contribution will come from the larger resonant mass peak of the heavier Higgs boson. The situation becomes increasingly complex when, in a given set of processes for a hypothesis, some have intermediate resonances while others do not. Nevertheless, such permutation symmetric arguments could effectively guide architecture design for cascade decays. 

One must, however, be cautious against the limitations of the narrow-width approximation~\cite{Berdine:2007uv}. The important takeaway message is that smaller group invariant approximations are not as overly constrained as larger ones: the smallest fibres of smaller group symmetries can become enlarged to those demanded by the larger one during training, but those of larger group invariant functions can not become smaller.  Therefore, for the case of observationally indistinguishable particles, the restriction to a smaller permutation symmetry does not induce any additional restrictions beyond the ones dictated by measurements.  Enlarging the symmetry in the case of distinguishable particles at reconstruction, like oppositely charged leptons, imposes the restrictions of NWA on the feature extraction even when the input data may contain effects beyond the NWA. An invariant graph readout over oppositely charged leptons, therefore, restricts the network to effects within the narrow width approximation in the case of resonant decays. To combine different processes into hypotheses, we would choose a permutation symmetry shared by all constituent processes. 

\subsection{Longitudinal Boost Equivariant Message Passing Neural Network}
\label{sec:longi_mpnn} 
Let us now construct an equivariant architecture looking into longitudinal boost equivariant quantities for a given final state $\mathbf{E}$.  While the same can be done within the formalism of refs.~\cite{pmlr-v119-bogatskiy20a,Bogatskiy:2022hub} or that of ref.~\cite{Spinner:2024hjm}, we choose the invariant theoretic formalism of refs.~\cite{pmlr-v139-satorras21a,villar2021scalars,Gong:2022lye,Li:2022xfc}, where one builds invariants and equivariant functions out of the basis of $\binom{n}{2}$ combinatorial dot products. Before going into detail, let us clarify the nature of the Lorentz group and its appropriate little groups concerning the fibre structures discussed above to guide the mathematical form of the architecture. 

Since we are eventually interested in invariant quantities, the graph readout should only propagate the invariant information. Within such an architecture, the feature extraction module by design has the smallest fibres of an invariant function, and one may erroneously conclude that intermediate equivariant updates are unimportant. However, the utility of function compositions (i.e. depth) in a neural network is to precisely induce successive topological changes in the data as evidenced in various studies~\cite{6697897,JMLR:v21:20-345}. Therefore, one cannot \textit{a priori} conclude that an invariant message passing update which induces larger minimal fibres of invariance from the beginning will behave the same as an equivariant update even though there is an invariant stage as one goes deeper in either network. Now, the equivariant updates of the longitudinal components $(p_z,E)$, already take care of the O(2) symmetry along the $z$-axis since it is the little group of the longitudinal boost action of the full Lorentz action, i.e. the longitudinally equivariant update of $(p_z,E)$ alone, make the fibres consists of rotations along the $z$-axis from the start. If one has a covariant expression of the complete four-vector update 
\begin{equation*}
 \mathbf{p}'_{\mu,i}=\mathbf{p}_{\mu,i}+\sum_j \mathbf{p}_{\mu,j}\Phi(\mathbf{p}_1,\mathbf{p}_2,....)\quad,
\end{equation*} $\Phi$ being a longitudinal boost invariant function, the transverse components will respect the vector action of the O(2) rotations around the $z$-axis, and hence be able to capture the equivariant information of the rotation. This is because in the $4\times 4$ matrix representation, longitudinal boosts and rotations along $z$-axis commute, i.e. we can break down the four-vector space as a direct sum of transverse and longitudinal components $p_\mu=(p_x,p_y)\oplus (p_z,E)$. However, in our final experiments, we only updated the longitudinal components and kept the O(2) invariant fibres from the beginning, as we did not find any additional performance gain.
We find that a scalar-only update performs just as well as the scalar-vector update for both the resonant and non-resonant di-Higgs searches.\footnote{There may be a relative difference in performance if one runs a hyperparameter scan which was not conducted for our case.}    
 
At the $(l+1)$-th stage of message passing, $l\geq 0$, let
 $\mathbf{\tilde{h}}^{(l)}_i$, $\mathbf{\tilde{e}}^{(l)}_{ij}$ be Lorentz scalar node-representation and edge representations, respectively.   Similarly, let $\mathbf{h}^{(l)}_i$ and $\mathbf{e}^{(l)}_{ij}$ be longitudinal boost invariant representations.
 With $\mathbf{\tilde{x}}^{(0)}_i=(p_x,p_y)$ and  $\mathbf{x}^{(l)}_i=(p^{(l)}_z,E^{(l)})_i$, the transverse and longitudinal components of the covariant four-vector $\mathbf{p}^{(l)}_i=(p_x,p_y,p^{(l)}_z,E^{(l)})_i$ we have 
 \begin{equation} 
 	\label{eq:four_vec_decomp}  \mathbf{p}^{(l)}_i=\mathbf{\tilde{x}}^{(0)}_i\oplus\mathbf{x}^{(l)}_i\quad.
 	\end{equation}  
 	Since all invariants of the Lorentz group are longitudinal boost invariant, let $\mathbf{\bar{h}}^{(l)}_{ij}$ and $\mathbf{\bar{e}}^{(l)}_{ij}$ be longitudinal boost invariant quantities which are not fully Lorentz invariant so that we have $\mathbf{h}^{(l)}_i=\mathbf{\bar{h}}_i\oplus\mathbf{\tilde{h}}_i$ and  $\mathbf{e}^{(l)}_{ij}=\mathbf{\bar{e}}^{(l)}_{ij}\oplus\mathbf{\tilde{e}}^{(l)}_{ij}$.  
The transverse component $\mathbf{\tilde{x}}^{(l)}_i$ being longitudinal boost invariant can be included in $\mathbf{\bar{h}}^{(l)}_i$, if one chooses only to update the longitudinal components but must be left out if we want an O(2) equivariant update of the transverse components.

With the notations clarified and abbreviating $\mathbf{p}^{(l)}_i+\mathbf{p}^{(l)}_j=\mathbf{p}^{(l)}_{ij}$, we can construct a longitudinal  equivariant message passing operation which updates  $\mathbf{h}^{(l)}_i$ and $\mathbf{x}^{(l)}_i$ as
\begin{equation}
	\label{eq:longi_lorentz}
	\begin{split}			
		\mathbf{m}^{(l+1)}_{ij}&=\Phi^{(l+1)}_e(\mathbf{h}^{(l)}_i,\mathbf{h}^{(l)}_j,\mathbf{e}^{(l)}_{ij},|\mathbf{p}^{(l)}_{ij}|_{(1,2)}^2,\langle\mathbf{p}^{(l)}_i,\mathbf{p}^{(l)}_j\rangle_{(1,2)},|\mathbf{p}^{(l)}_{ij}|^2,\langle\mathbf{p}^{(l)}_i,\mathbf{p}^{(l)}_j\rangle)\quad,\\	
		\mathbf{x}_i^{(l+1)}&=\mathbf{x}^{(l)}_i+\sum_{j\in\mathcal{N}(i)} \;
		\mathbf{x}^{(l)}_j\;\Phi^{(l+1)}_x(\mathbf{m}^{(l+1)}_{ij}) \quad, \\			 
		\mathbf{m}^{(l+1)}_i&=\frac{1}{|\mathcal{N}(i)|}\sum_{j\in\mathcal{N}(i)} \;\mathbf{m}^{(l+1)}_{ij}\quad,\\
		\mathbf{h}^{(l+1)}_i&=\Phi^{(l+1)}_h(\mathbf{h}^{(l)}_i,\mathbf{m}^{(l+1)}_i)  \quad. \\	
	\end{split}
\end{equation} 
The functions $\Phi_e^{(l+1)}$, $\Phi_x^{(l+1)}$, and $\Phi_h^{(l+1)}$ are multi-layer perceptrons (MLP), with $\Phi_x^{(l+1)}$ giving a one-dimensional weight after a sigmoid activation on the final layer. While we have included a node-update function $\Phi^{(l+1)}_h$, we have used $\mathbf{m}_i^{(l+1)}=\mathbf{h}_i^{(l+1)}$ in our experiments as there was no relative difference in the performance.

\section{Illustrative example: Di-Higgs to four bottom jets} 
\label{sec:dihiggs}
We employ the challenging but important di-Higgs search in the four bottom decay channels to test the methodology developed in the previous section. A recent work~\cite{Chiang:2024pho} utilising Symmetry Preserving Attention Networks (\textsc{Spa-Net})~\cite{Fenton:2020woz,Shmakov:2021qdz,Fenton:2023ikr} achieved state-of-the-art performance in the resonant and non-resonant production channel of the two Higgs boson, where in the former, there is an additional BSM heavy scalar boson which then resonantly decays to the two SM Higgs. As discussed above, while the final signatures are the same for both signals, they have inherently different approximate permutation symmetries. Moreover, we use the same data made public~\cite{hsieh_2024_10952296} by the authors, with the only essential difference coming from the network analysis. 
\subsection{Dataset description} 
We highlight the important elements of the utilised dataset. Parton level events were generated using \texttt{MadGraph5\_aMC@NLO (v3.3.1)}~\cite{Alwall:2014hca} at $E_{cm}=13$ TeV, which were showered and hadronised with \texttt{Pythia8.306}~\cite{Bierlich:2022pfr}. All stable hadrons went through a detector simulation in \texttt{Delphes (v3.5.0)}~\cite{deFavereau:2013fsa}. In the object reconstruction, \texttt{FastJet (v3.3.4)}~\cite{Cacciari:2011ma} was used to cluster anti-$k_t$~\cite{Cacciari:2008gp} jets with radius $R=0.4$ and transverse momentum $p_T\geq20$ GeV.    For the resonant analysis, the b-tagging efficiencies were modified at the 70\% working point of the ATLAS MV2c10 b-tagger~\cite{ATLAS:2016gsw,ATLAS:2015thz}. In contrast, the non-resonant case was modified to the 77\% working point of ATLAS DL1r tagger~\cite{ATLAS:2022qxm}. Selected events contain at least four b-tagged jets with $p_T>40$ GeV and $|\eta|<2.5$.  We refer interested readers to ref.~\cite{Chiang:2024pho} for more data generation and baseline selection details. 
 \subsection{Preprocessing and data representation} 
 In each event, we use the four hardest b-tagged jets to form the two Higgs candidates using the $\Delta R+\min D_{hh}$ cut-based pairing motivated by the ATLAS analysis~\cite{ATLAS:2018rnh} also utilised in the cut-based pairing in the dense neural network input in ref.~\cite{Chiang:2024pho} with a minor difference. For the $\Delta R$ requirement, defining the candidate with leading $p_T$ as $h_1$ and the other as $h_2$, one considers the cut 
 \begin{equation}
 	\begin{split}
 		\frac{360~\text{GeV}}{m_{4j}} - 0.5 &< \Delta R^{h_1}_{bb} < \frac{653~\text{GeV}}{m_{4j}} + 0.475\\
 		\frac{235~\text{GeV}}{m_{4j}} &< \Delta R^{h_2}_{bb} < \frac{875~\text{GeV}}{m_{4j}} + 0.35
 	\end{split}
 	\end{equation}
 	if $m_{4b} < 1250$ GeV over the possible bottom jet pairings and 
 	\begin{equation} 
 	\begin{split}
 		0&< \Delta R^{h_1}_{bb} < 1\\
 		0 &< \Delta R^{h_2}_{bb} < 1
 	\end{split}
 	\end{equation}
 	if $m_{4b} > 1250$ GeV. For those events having more than one instance of the partitions passing the above requirements, the one with the minimum $D_{hh}$ defined as 
 	\begin{equation}
 		D_{hh}=\left|m_{h_1}-\frac{120}{110}m_{h_2}\right|\left( 1+\frac{120^2}{110^2}\right)^{-1/2}\quad,
 	\end{equation}
 	is chosen to be the Higgs candidate. In contrast to the above-mentioned analyses, we do not drop the event if no partitions pass the $\Delta R$ criterion and use the minimum $D_{hh}$ pair over all possible pairs in such events to specify the possible Higgs candidates. These possible Higgs candidates segregate the four bottom jets into two classes of reconstructed objects: $H_1$ and $H_2$. Any other jet in the reconstructed event, including additional b-jets, is classified under a single jet class $J$.  
  
 	After segregating the reconstructed jets into the three classes, we construct a complete graph with edges connecting all distinct objects, i.e. without self-loops.  The input node representations consist of the Lorentz four-vector $\mathbf{p}^{(0)}_i$, and the longitudinal scalar node representation\footnote{A statistically negligible amount of events in the dataset had jets with zero mass and were excluded from all numerical analyses.}
  $$\mathbf{h}^{(0)}_i=(\phi_i,\log p^t_i,\log m^t_i, b_i,\log m_i)\quad,$$
  consists of the jets' azimuthal angle $\phi_i$ , transverse momentum $p^t_i$ , transverse mass $m^t_i=\sqrt{E_i^2-p_z^2}$, b-tagging information $b_i\in\{-1,1\}$, and mass $m_i$. We set $b_i=1$ for a b-tagged jet. Each edge has a longitudinal scalar edge-representation $$\mathbf{e}^{(0)}_{ij}=(p^t_{ij},\log (p^t_i\,p^t_j),\Delta \eta_{ij},\Delta \phi_{ij},\Delta R_{ij})\quad,$$
where $p^t_{ij}$ is the transverse momentum of $\mathbf{p}_i+\mathbf{p}_j$, $\Delta \eta_{ij}$ is the difference in pseudorapidity, $\Delta \phi_{ij}$ the azimuthal separation and $\Delta R_{ij}=\sqrt{\Delta \eta_{ij}^2+\Delta \phi_{ij}^2}$. For the O(1,3) network, we consider only the fully Lorentz invariant\footnote{Strictly speaking, the b-tagging information $b_i$ being dependent on reconstruction is not Lorentz invariant. However, as is usually done in most applications, we assume that it reflects the true flavour of the underlying primaeval parton.} node features, $\mathbf{\tilde{h}}^{(0)}_i=(b_i,\log m_i)$ and do not supply any additional edge feature since the message passing operation automatically evaluates the relevant edge invariants.\footnote{While one could argue that the Lorentz invariant model has less information supplied, this is a mandatory requirement: larger group invariances assume that information contained within the separate orbits of its proper sub-groups are the same and therefore not relevant.} The classes $H_1$ and $H_2$ undergo a mean global mean readout either separately (for $S_2\times S_2$ group) or together (for $S_4$ group), along with any additional jets which are uniformly given a weight of $w_{\alpha\leftarrow J}=0.001$ for $\alpha\in\{H_1,H_2,H_1\cup H_2\}$.

\subsection{Network Analysis}
Looking into graph-based architectures, a segregation of the reconstructed objects allows for a heterogeneous graph message-passing operation, which preserves all symmetries of the likelihood ratio.  On the other hand, we want to learn the kinematic correlations between the different classes efficiently.  This can be achieved in the heterogeneous set-up with multiple copies of learnable functions for the node and edge type combinatorics. Since this scales factorially, if we consider edge directions, we choose the simpler homogeneous message-passing operation with the learnable functions shared between all nodes and edges. All network analyses uses \textsc{Pytorch-Geometric (v2.5.0)}~\cite{Fey/Lenssen/2019} and \textsc{PyTorch (v2.0.0)}~\cite{10.5555/3454287.3455008}. 

We consider three base architectures with different message-passing heads:
\begin{enumerate} 
\item O(1,1)-S : a scalar-only longitudinal boost invariant message passing head. This is essentially an $\texttt{EdgeConv}$~\cite{10.1145/3326362} network that takes $\mathbf{h}^{(0)}_i$ and $\mathbf{e}^{(0)}_{ij}$  as inputs. 
\item O(1,1)-SV : a scalar-and-vector update longitudinal boost equivariant message passing head
\item O(1,3) : a Lorentz Group Equivariant Block~\cite{Gong:2022lye} modified so that $\Phi_e$ takes $|\mathbf{p}^{(l)}_{i}+\mathbf{p}^{(l)}_j|^2$ instead of their choice of momentum difference squared inputs and no\footnote{We did not find any noticeable performance difference with the addition of $\Phi_h$.} $\Phi_h$.  
\end{enumerate}

Similar to ref.~\cite{Gong:2022lye}, all inner products and norms go through the function $R(x)=\text{sign}(x)\log(|x|+1)$, so that the gradient descent is stable for the non-compact metric signature. Each model has a wide variant of 256, 128, and 64 updated scalar-node dimensions and a narrow variant of 64, 32, and 16 updated scalar-node representations. All MLPs have two hidden layers with the same dimensions as their respective scalar update dimensions with \texttt{ReLU} activation in the hidden layers. The output layers have \texttt{Linear} activations except for $\Phi^{(l)}_x$, which has a \texttt{Sigmoid} activation function. The message functions $\Phi^{(l)}_e$ in O(1,1)-SV and O(1,3) models take additional edge scalar edge features evaluated at each stage $l$. The O(1,1)-S model consists of only the $\Phi_e$ function in each stage, which takes the scalar representation $\mathbf{h}^{(l)}_i$ and $\mathbf{h}^{(l)}_j$ without any additional edge features beyond the initial input operation. Additionally, $\Phi^{(l)}_e$ in  O(1,1)-S and O(1,1)-SV evaluates the \texttt{EdgeConv} input $\mathbf{h}^{(l)}\oplus \mathbf{h}^{(l)}_j-\mathbf{h}^{(l)}_i$ from the scalar node representations in each stage $l$ of the message passing head. All three base architectures have a mean scalar node readout. For all networks, the updated scalar node-representations $\mathbf{h}^{(l)}_i$, for $l>0$ undergoes a global mean readout which is either $S_2\times S_2$ invariant or $S_4$ invariant depending on the discrete symmetry of the network. Consequently, the final message passing operation for O(1,1)-SV and O(1,3) does not have a vector update operation.  The respective node representations and the permutation symmetry determine the inputs to the classifier head. The classifier MLP has two hidden layers of 64 nodes and \texttt{ReLU} activation for the wide message passing head, while the ones with narrow message passing heads have 32 nodes instead. With a single logit output, the networks are trained with \texttt{torch.nn.functional.binary\_cross\_entropy\_with\_logits} loss function.    

Counting the wide and narrow variants of the message-passing heads and the global readout symmetry, we have four network architectures for each base architecture. These four instances are trained on two training sizes for the resonant and non-resonant cases: the full dataset and a reduced set containing 100k samples for the resonant case and 10k for the non-resonant case. We use the test dataset as the validation set during training and utilise the complete training dataset for the first case.\footnote{The difference of 50k and 9k training samples from ref.~\cite{Chiang:2024pho} for the resonant and non-resonant cases, respectively, is not a major difference for the quoted results as network performance generally scales logarithmically with training size. Concretely, the smallest O(1,1)-S network with an $S_4$ invariant global readout with 22k trainable parameters reached an AUC of 0.9632 on the test dataset with 600k training samples on the resonant signal dataset.} 
Networks in each experiment are trained ten times after random weight initialisation with the \texttt{Adam}~\cite{DBLP:journals/corr/KingmaB14} optimiser with an initial learning rate of 0.001 and a batch size of 128 samples-per-batch. A decay-on-plateau condition decays the learning rate if the validation loss has not improved for five epochs by a factor of 0.1 until it reaches $10^{-8}$. The training runs for a maximum of one hundred epochs and is stopped if the validation loss has not decreased for twenty epochs.  
  
\subsection{Performance }
\begin{table}
\begin{center}
\resizebox{\textwidth}{!}{
\begin{tabular}{ccccl}
\toprule
Arch. & Signal & Disc. Sym. & Num. Param. & AUC  \\
\midrule
\multirow{2}{*}{O(1,1)-S}   & Resonant & $S_4$ & 293k & 0.9652$\pm$0.0002 \\
&Non-resonant &$S_4$ & 22k & 0.9165$\pm$0.0005  \\
\hline 
\multirow{2}{*}{O(1,1)-SV}& Resonant & $S_4$ & 458k & 0.9653$\pm$0.0001  \\
& Non-resonant& $S_4$ & 33k & 0.9169$\pm$0.0009  \\
\hline 
 \multirow{2}{*}{O(1,3)} & Resonant & $S_2 \times S_2$ & 743k & 0.9550$\pm$0.0016  \\
& Non-resonant& $S_2 \times S_2$ & 743k & 0.9000$\pm$0.0009 \\
\midrule 
\multirow{2}{*}{\textsc{Spa-Net} (ref.~\cite{Chiang:2024pho})}& Resonant & $S_n$ & 37.9M& $0.961 \pm 0.001$ \\
&Non-resonant & $S_n$ & 541k &$0.911 \pm 0.001$ \\
\bottomrule

\end{tabular}
}
\caption{The best AUC score out of all experiments conducted for each base architecture on the full dataset of each signal scenario. The mean and standard deviation is taken over ten training instances from random weight initialisation. For comparison, we show the relevant figures for \textsc{Spa-Net}.}
\label{tab:best_exps}
\end{center}
\end{table}

  For each training experiment, we evaluate the area under the curve (AUC) under the receiver operator characteristics (ROC) curve over each training instance from which we form various summary statistics of the performance metrics. Here, we report the main findings while all results are tabulated in appendix~\ref{app:add_res}. The best AUC score for each base architecture over the two datasets, along with the details of the specific architecture, is shown in table~\ref{tab:best_exps}. The figures of \textsc{Spa-Net} from \cite{Chiang:2024pho} are also shown for comparison. Lorentz invariant classification fares poorly in either scenario compared to O(1,1)-S and O(1,1)-SV and can not match the \textsc{Spa-Net} results, which do not assume any continuous group equivariance. The correct continuous group symmetric design of O(1,1)-S and O(1,1)-SV outperforms \textsc{Spa-Net} with an order of magnitude reduction in trainable parameters. This is all the more impressive considering that the numerical experiments for the \textsc{Spa-Net} based analysis conducted a hyperparameter scan. Additionally, the low parameter-size networks perform nominally better for the non-resonant scenario than the highly parametrised ones (see table~\ref{tab:nres} in appendix~\ref{app:add_res}). This could be due to the lower statistics of the training data in the non-resonant dataset, where a larger model size performs better with more training statistics. On the contrary, the incorrect invariance in O(1,3) has the wider network performing better than the smaller network, even with the limited training statistics of the non-resonant training dataset. This may be 
due to the assumption of an incorrect exact invariance in the domain and the presence of noise in the data, which requires more model flexibility to circumvent the exact symmetric design of the architecture. This intuition could also help explain the better performance of the smaller $S_2\times S_2$ permutation symmetry for O(1,3) for either signal scenario, where the larger $S_4$ symmetry comparatively over-constrains the fibres of the target function.

\begin{figure}
\centering 
	\includegraphics[scale=0.20]{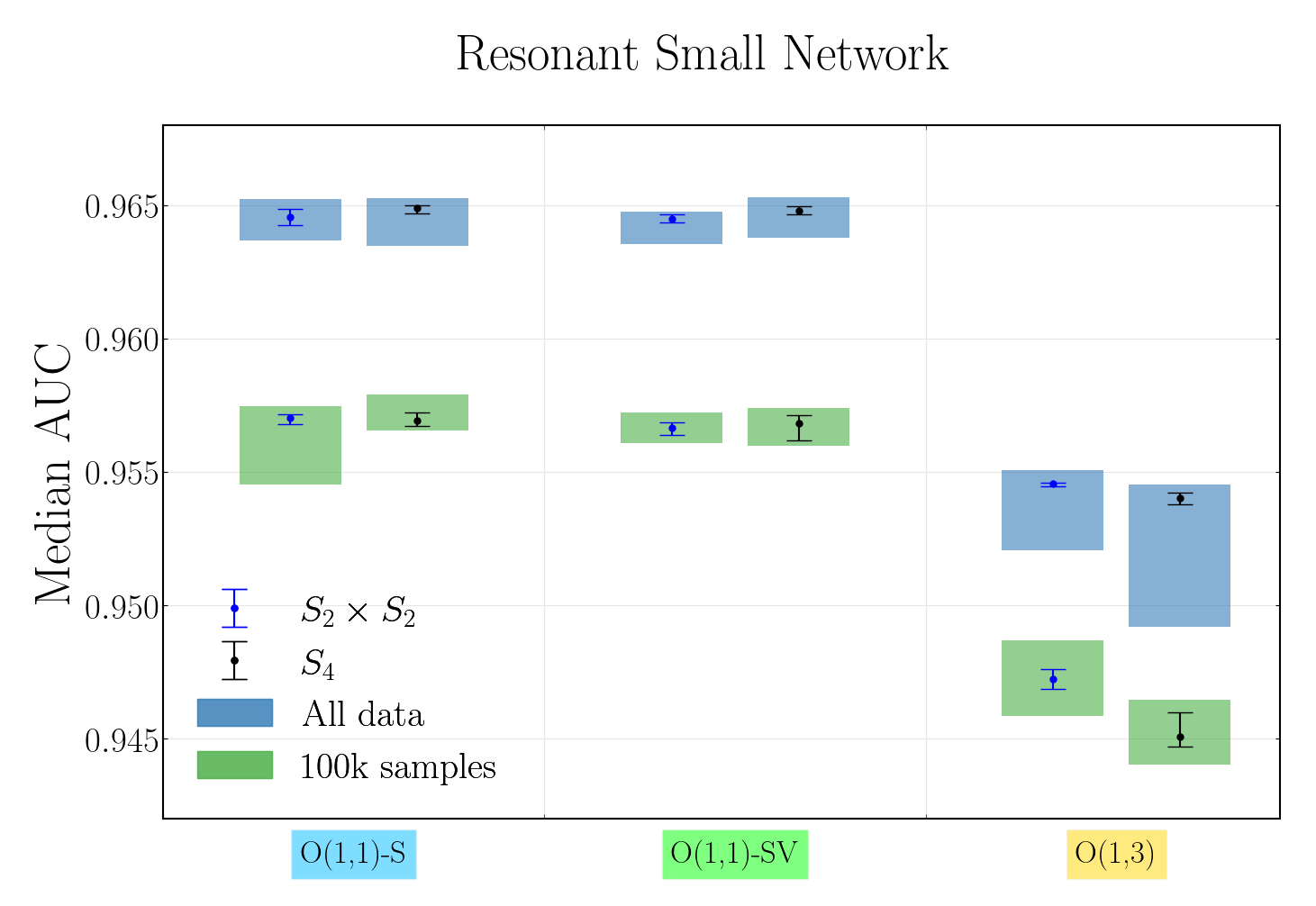}
	\includegraphics[scale=0.20]{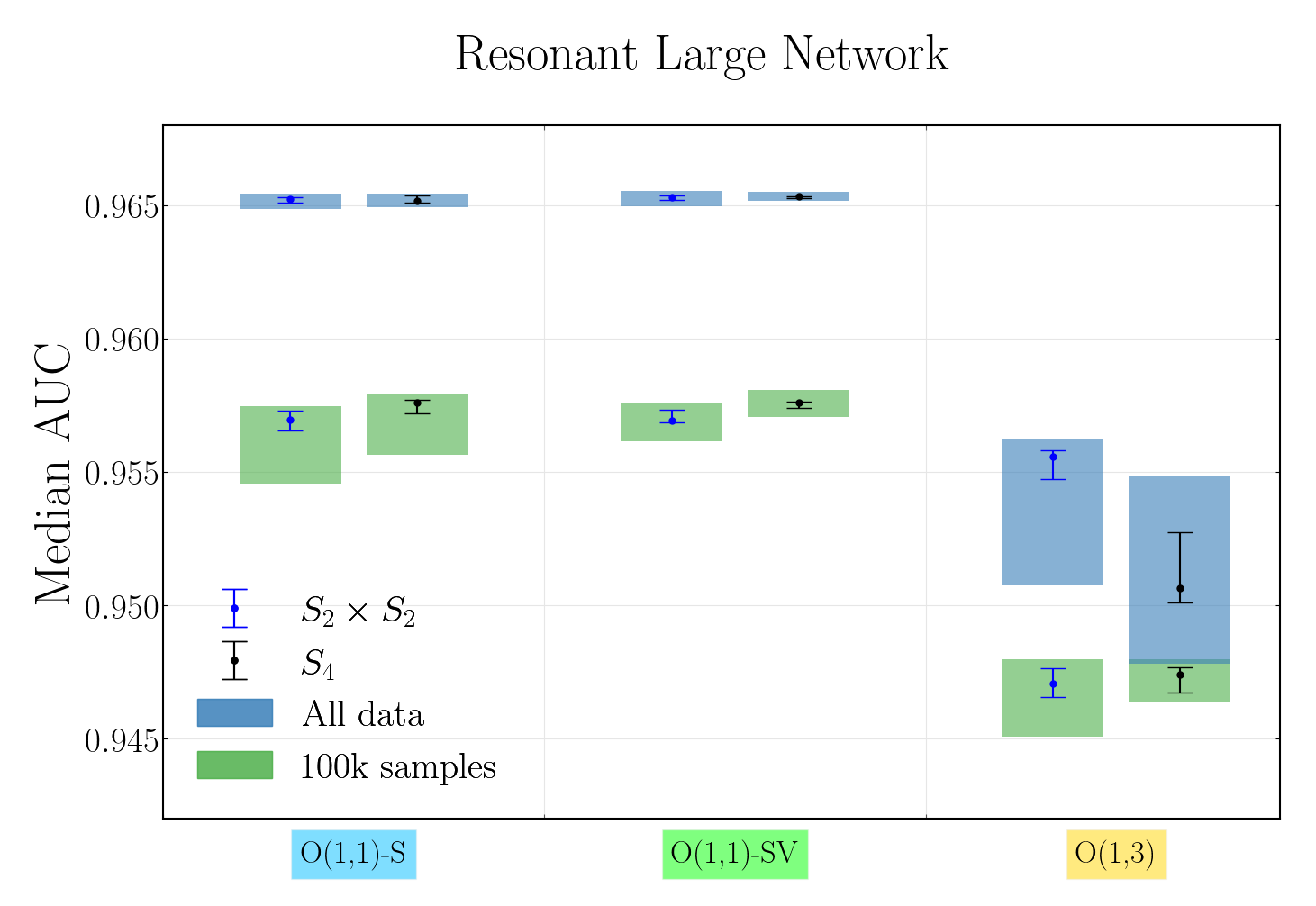}
	\caption{Median AUCs for the resonant signal for all architectures and training sizes. The shaded region shows the range, and the error bars denote the lower and upper quartiles. }
 \label{fig:res_auc} 
\end{figure}

\begin{figure}
\centering 
	\includegraphics[scale=0.20]{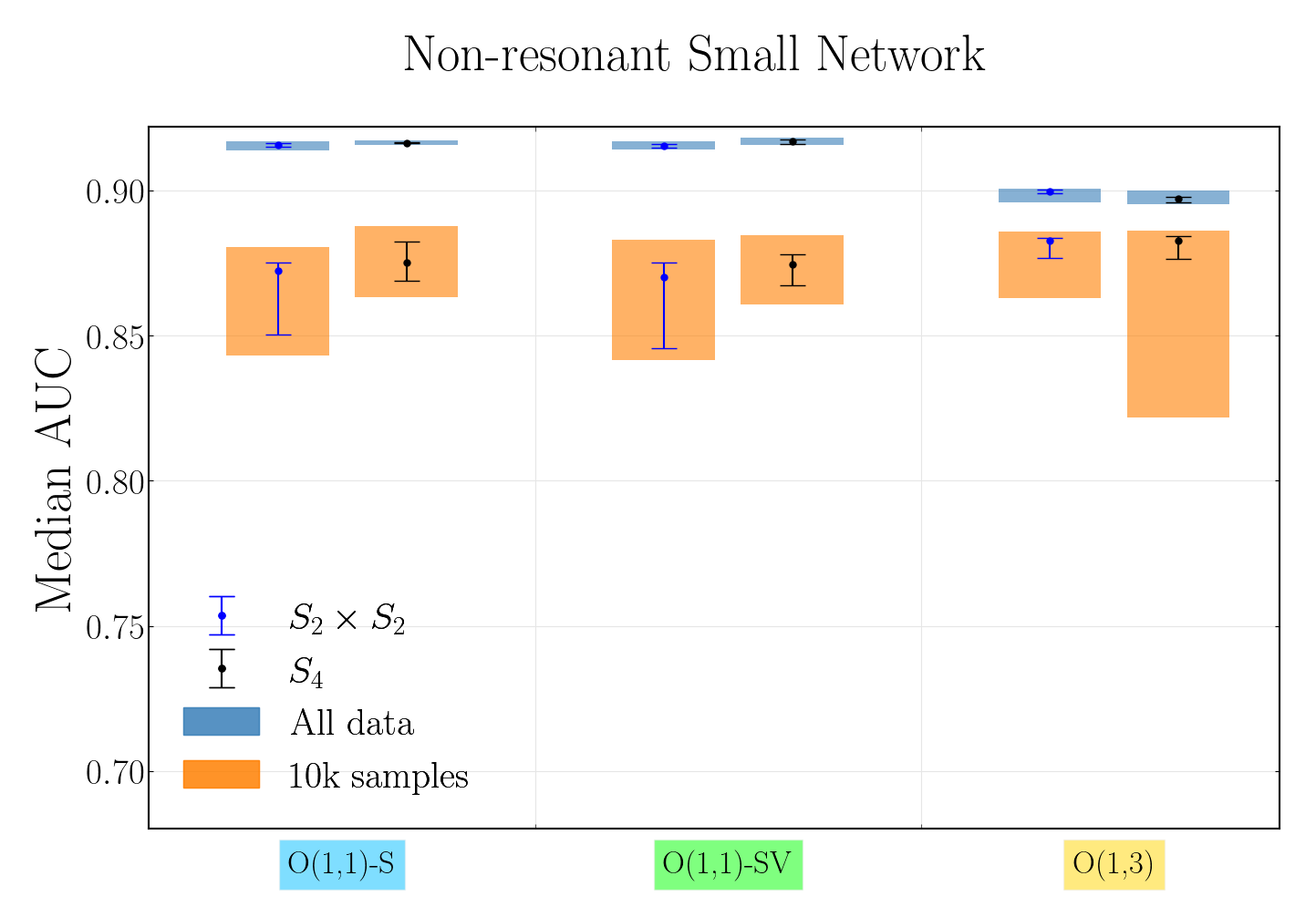}
	\includegraphics[scale=0.20]{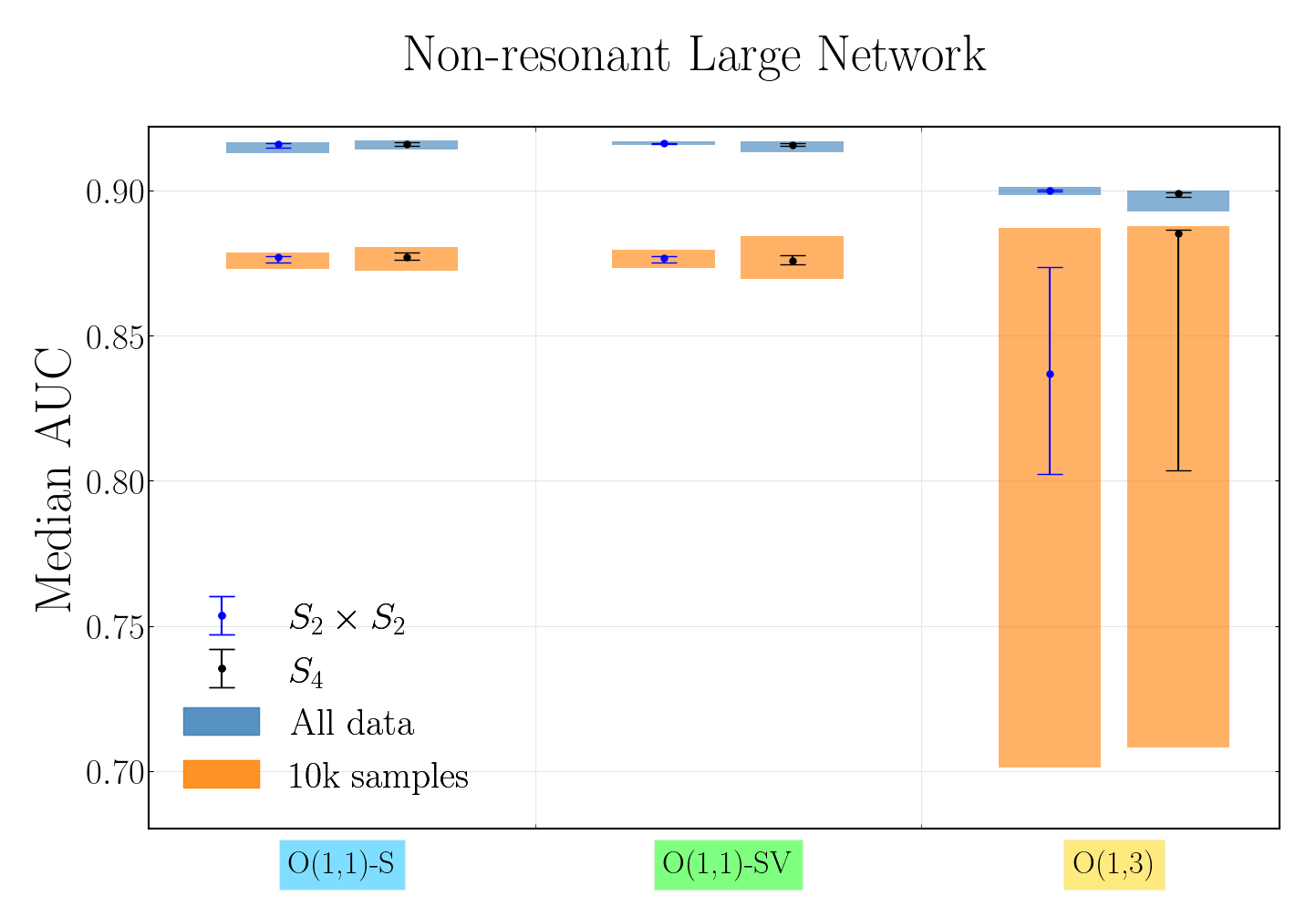}
	\caption{Median AUCs for the non-resonant signal for all architectures and training sizes. The shaded region shows the range, and the error bars denote the lower and upper quartiles. }
	\label{fig:nres_auc} 
\end{figure}

The median of the AUC and its lower and upper quartiles as error bars for each base architecture and training data size are plotted in fig.~\ref{fig:res_auc}  for the resonant scenario and fig.~\ref{fig:nres_auc} for the non-resonant one.  For both signal scenarios, the choice of the discrete symmetry group has nominal differences in the median values for the O(1,1)-S and O(1,1)-SV architectures that have the correct continuous invariance.   The smaller networks have a larger range for both signal scenarios, suggesting a trade-off between training stability and parameter complexity. 
Similarly, the low training statistics cases have larger ranges for the correct continuous equivariance than the full dataset training. 
The situation is mostly reversed in the case of O(1,3) networks, where the continuous symmetry is incorrect. As seen above, the smaller group $S_2\times S_2$ has better overall median AUCs than the larger $S_4$ symmetric readouts, barring the non-resonant large-network experiment in the low training sample scenario. However, in this situation, both networks have very erratic behaviour over the ten training instances, as can be seen by the large range and extreme position of the median values.  Interestingly, all networks with the correct continuous symmetry, regardless of the network size and discrete symmetry group, outperform \textsc{Spa-Net} on the full dataset.

\section{Conclusions} 
In this work, we have established a novel connection between the Matrix-Element Method (MEM) and equivariant neural network architecture design, demonstrating how MEM-inspired symmetries can guide the development of deep learning models for high-energy physics analysis. By incorporating a suitable subgroup of the physical Lorentz and permutation invariances directly into the architecture, we have shown that neural networks can achieve improved performance in classification tasks while maintaining lower parameter complexity.

 Our approach uses the inherent symmetry properties embedded in fixed-order differential cross-sections and exploits the optimality of group-equivariant functions for binary classification. We demonstrated that designing neural networks with MEM-inspired equivariant updates results in architectures that better capture the kinematic correlations of events, especially for complex final states, such as di-Higgs production decaying to four bottom jets. The longitudinal boost-equivariant message-passing network proposed in this work provides a concrete example of how these principles can be applied to practical physics problems, yielding state-of-the-art performance on benchmark datasets. Moreover, the analysis reveals that smaller group invariance approximations can effectively generalise to larger symmetries during training, while larger group invariance constraints might overlook subtle details in the data.

Our findings open several avenues for future research. First, extending these principles to higher-dimensional final states and more complex processes, such as multi-jet events or processes with additional intermediate resonances, could further elucidate the benefits of MEM-inspired equivariant architectures. Additionally, integrating such symmetric architecture designs with other advanced deep learning techniques, such as transformers or attention mechanisms, could offer even more powerful tools for particle physics analysis. Furthermore, applying this framework to multi-class classification problems in physics searches, where different classes exhibit distinct symmetry properties, could improve LHC's sensitivity to new physics.

Thus, this study demonstrates that integrating MEM with equivariant deep learning techniques can significantly enhance neural networks' capabilities in high-energy physics. By grounding the architecture design in physical principles, we can improve model interpretability, reduce computational requirements, and potentially uncover new physics beyond the Standard Model.

\section*{Acknowledgements}
This work is carried out under the support of STFC under grant ST/X003167/1. We thank Matteo Marcoli for the helpful discussions during the project. We also thank the authors of ref~\cite{Chiang:2024pho} for making their dataset publicly available.

\appendix

\section{Additional results of network analysis}
This appendix shows the results of all numerical experiments conducted on the resonant and non-resonant datasets. Including the AUC, we show the $R_{30}$ and $R_{50}$ metrics defined as the inverse of the background acceptance (false positive rate) at 30 and 50 per cent signal acceptances (true positive rate), respectively. These are shown for the resonant and non-resonant signals in tables~\ref{tab:res} and \ref{tab:nres}, respectively. One can confirm that the correct continuous group symmetries, regardless of the network size and permutation symmetries, outperform \textsc{Spa-Net} on the full dataset.   

\label{app:add_res}
\begin{table} 
\begin{center} 
\resizebox{\textwidth}{!}{
\begin{tabular}{cccclll}
\toprule
Arch. & Train. Size & Disc. Sym. & Num. Param. & AUC & $R_{30}$ & $R_{50}$ \\
\midrule
\multirow{8}{*}{O(1,1)-S} & \multirow{4}{*}{All} & \multirow{2}{*}{$S_4$} & 293k & 0.9652$\pm$0.0002 & 2135$\pm$303 & 375$\pm$14 \\
&  &  & 22k & 0.9647$\pm$0.0005 & 2000$\pm$139 & 362$\pm$14 \\
\cline{3-7}
 &  & \multirow{2}{*}{$S_2 \times S_2$} & 322k & 0.9652$\pm$0.0002 & 2037$\pm$269 & 363$\pm$18 \\
 &  &  & 26k & 0.9646$\pm$0.0005 & 2000$\pm$227 & 361$\pm$7 \\
 
\cline{2-7}
 & \multirow{4}{*}{100k} & \multirow{2}{*}{$S_4$} & 293k & 0.9572$\pm$0.0008 & 1274$\pm$129 & 257$\pm$22 \\
  &  & & 22k & 0.9570$\pm$0.0004 & 1357$\pm$115 & 266$\pm$12 \\
\cline{3-7}
 &  & \multirow{2}{*}{$S_2 \times S_2$} & 322k & 0.9567$\pm$0.0009 & 1154$\pm$141 & 259$\pm$17 \\

 &  & & 26k & 0.9567$\pm$0.0009 & 1287$\pm$160 & 262$\pm$11 \\

\midrule
\multirow{8}{*}{O(1,1)-SV}& \multirow{4}{*}{All} & \multirow{2}{*}{$S_4$} & 458k & 0.9653$\pm$0.0001 & 2137$\pm$260 & 370$\pm$12 \\
 &  &  & 33k & 0.9648$\pm$0.0004 & 2116$\pm$321 & 356$\pm$15 \\
\cline{3-7}
 &  & \multirow{2}{*}{$S_2 \times S_2$} & 487k & 0.9653$\pm$0.0002 & 2064$\pm$291 & 357$\pm$10 \\
 &  & & 36k & 0.9644$\pm$0.0003 & 2060$\pm$171 & 367$\pm$9 \\

\cline{2-7} 
& \multirow{4}{*}{100k} & \multirow{2}{*}{$S_4$} & 458k & 0.9575$\pm$0.0003 & 1281$\pm$161 & 259$\pm$13 \\
&  & & 33k & 0.9567$\pm$0.0005 & 1394$\pm$123 & 263$\pm$10 \\
 \cline{3-7} 
 &  & \multirow{2}{*}{$S_2 \times S_2$} & 487k & 0.9570$\pm$0.0004 & 1149$\pm$99 & 262$\pm$10 \\
&  & & 36k & 0.9567$\pm$0.0004 & 1343$\pm$103 & 259$\pm$13 \\

\midrule
\multirow{8}{*}{O(1,3)} & \multirow{4}{*}{All} & \multirow{2}{*}{$S_4$} & 715k & 0.9512$\pm$0.0024 & 784$\pm$100 & 156$\pm$14 \\
 &  &  & 48k & 0.9536$\pm$0.0016 & 886$\pm$82 & 169$\pm$8 \\
\cline{3-7} 
 &  & \multirow{2}{*}{$S_2 \times S_2$} & 743k & 0.9550$\pm$0.0016 & 865$\pm$67 & 180$\pm$7 \\
 &  & & 52k & 0.9542$\pm$0.0011 & 864$\pm$66 & 171$\pm$7 \\

\cline{2-7}
 & \multirow{4}{*}{100k} & \multirow{2}{*}{$S_4$} & 715k & 0.9472$\pm$0.0006 & 643$\pm$30 & 141$\pm$5 \\
&  &  & 48k & 0.9453$\pm$0.0008 & 599$\pm$33 & 135$\pm$3 \\
\cline{3-7}
&  & \multirow{2}{*}{$S_2\times S_2$} & 743k & 0.9470$\pm$0.0009 & 618$\pm$50 & 137$\pm$4 \\
&  &  & 52k & 0.9473$\pm$0.0008 & 630$\pm$27 & 141$\pm$4 \\

\bottomrule

\end{tabular}
}
\caption{The mean AUC, $R_{30}$, and $R_{50}$ for all experiments on the resonant dataset.}
\label{tab:res}
\end{center}
\end{table}

\begin{table}
\begin{center} 
\resizebox{\textwidth}{!}{
\begin{tabular}{cccclll}
\toprule
Arch. & Train. Size & Disc. Sym. & Num. Param. & AUC & $R_{30}$ & $R_{50}$ \\
\midrule
\multirow{8}{*}{O(1,1)-S} & \multirow{4}{*}{All} & \multirow{2}{*}{$S_4$} & 293k & 0.9160$\pm$0.0009 & 236$\pm$20 & 49$\pm$2 \\
& & & 22k & 0.9165$\pm$0.0005 & 256$\pm$25 & 50$\pm$2 \\
\cline{3-7}
& & \multirow{2}{*}{$S_2 \times S_2$} & 322k & 0.9155$\pm$0.0013 & 231$\pm$21 & 48$\pm$2 \\
& & & 26k & 0.9158$\pm$0.001 & 248$\pm$14 & 49$\pm$3 \\
\cline{2-7} 
& \multirow{4}{*}{10k} & \multirow{2}{*}{$S_4$} & 293k & 0.8772$\pm$0.0023 & 82$\pm$6 & 22$\pm$1 \\
&  & & 22k & 0.8754$\pm$0.0087 & 83$\pm$19 & 22$\pm$4 \\
\cline{3-7}
& & \multirow{2}{*}{$S_2 \times S_2$} & 322k & 0.8764$\pm$0.0018 & 78$\pm$6 & 22$\pm$1 \\
& & & 26k & 0.8645$\pm$0.0148 & 67$\pm$16 & 18$\pm$4 \\
\midrule 
\multirow{8}{*}{O(1,1)-SV} & \multirow{4}{*}{All} & \multirow{2}{*}{$S_4$} & 458k & 0.9157$\pm$0.001 & 227$\pm$18 & 50$\pm$2 \\
&&& 33k & 0.9169$\pm$0.0009 & 272$\pm$29 & 50$\pm$1 \\
\cline{3-7} 
& & \multirow{2}{*}{$S_2 \times S_2$} & 487k & 0.9164$\pm$0.0004 & 238$\pm$28 & 50$\pm$2 \\
& & & 36k & 0.9155$\pm$0.0009 & 243$\pm$21 & 49$\pm$2 \\
\cline{2-7}
 & \multirow{4}{*}{10k} & \multirow{2}{*}{$S_4$} & 458k & 0.8765$\pm$0.004 & 81$\pm$9 & 22$\pm$1 \\
& & &  33k& 0.8729$\pm$0.0076 & 81$\pm$20 & 20$\pm$3 \\
\cline{4-7} 
& & \multirow{2}{*}{$S_2 \times S_2$} & 487k & 0.8766$\pm$0.0018 & 81$\pm$6 & 22$\pm$1 \\
& & & 36k & 0.863$\pm$0.0167 & 65$\pm$18 & 18$\pm$5 \\

\midrule
\multirow{8}{*}{O(1,3)} & \multirow{4}{*}{All} & \multirow{2}{*}{$S_4$} & 715k & 0.8981$\pm$0.0022 & 123$\pm$8 & 31$\pm$1 \\
 &  & & 48k & 0.8972$\pm$0.0014 & 125$\pm$4 & 30$\pm$1 \\
\cline{3-7}
& & \multirow{2}{*}{$S_2 \times S_2$} & 743k & 0.9000$\pm$0.0009 & 130$\pm$7 & 30$\pm$1 \\
 & & & 52k & 0.8995$\pm$0.0014 & 129$\pm$10 & 31$\pm$1 \\
\cline{2-7} 
 & \multirow{4}{*}{10k} & \multirow{2}{*}{$S_4$} & 715k & 0.8437$\pm$0.0662 & 82$\pm$47 & 19$\pm$9 \\
  &  &  & 48k & 0.8737$\pm$0.0209 & 92$\pm$26 & 22$\pm$5 \\
 \cline{3-7}
 & & \multirow{2}{*}{$S_2 \times S_2$} & 743k & 0.8219$\pm$0.0663 & 50$\pm$32 & 14$\pm$8 \\

 & &  & 52k & 0.8793$\pm$0.0075 & 93$\pm$15 & 23$\pm$3 \\
\bottomrule
\end{tabular}
}
\caption{The mean AUC, $R_{30}$, and $R_{50}$ for all experiments on the non-resonant dataset.}
\label{tab:nres}
\end{center}
\end{table}

\bibliographystyle{JHEP}
\bibliography{ref}
\end{document}